\documentclass[12pt]{article}

\usepackage{color,extarrows} 
\numberwithin{equation}{section}

\newcommand{\be}{\begin{equation}}
\newcommand{\ee}{\end{equation}}
\usepackage{amsmath}
\usepackage{amssymb}
\usepackage{latexsym}
\usepackage{epsfig}
\usepackage{multirow}
\usepackage{lscape}

\newcommand{\del}{\partial}

\newcommand{\bea}{\begin{eqnarray}}
\newcommand{\eea}{\end{eqnarray}}

\newcommand{\RRR}{{\hbox{\rm R\kern-2.35mm R}}}

\def\ZZZ{{\hbox{ Z\kern-1.6mm Z}}}

\def\a{\alpha}

\def\cG{{\cal G}}

\begin{document}

\begin{titlepage}
\rightline{September 2019}
\rightline{  Imperial-TP-2019-CH-07}
\begin{center}
\vskip 2.5cm
{\Large \bf {
The Doubled Geometry of Nilmanifold Reductions}}\\
\vskip 2.0cm
{\large {N.  Chaemjumrus and C.M. Hull  }}
\vskip 0.5cm
{\it {The Blackett Laboratory}}\\
{\it {Imperial College London}}\\
{\it {Prince Consort Road}}\\
{\it { London SW7 @AZ, U.K.}}\\

\vskip 2.5cm
{\bf Abstract}
\end{center}

\vskip 0.5cm

\noindent
\begin{narrower}
A  class of special holonomy spaces arise as nilmanifolds fibred over a line interval and are dual to intersecting brane solutions of string theory. Further dualities relate these to T-folds, exotic branes, essentially doubled spaces and spaces with R-flux. We develop the doubled geometry of these spaces, with the various duals arising as different slices of the doubled space.

\end{narrower}

\end{titlepage}

\newpage

\tableofcontents
\baselineskip=16pt
\section{Introduction}

Nilmanifolds are compact spaces obtained by quotienting a nilpotent Lie group by a discrete subgroup.  The 3-dimensional one obtained from the Heisenberg group is  referred to as the nilfold, and is T-dual to a 3-torus with $H$-flux. It is not a solution of string theory, but it appears as the fibre in a 4-dimensional hyperk\" ahler space which is a bundle over a line interval, and so can be incorporated into string theory in this way~\cite{Hull:1998vy, Lavrinenko:1997qa,Gibbons:1998ie}.
 This case and its duals were explored in detail in \cite{Chaemjumrus:2019ipx}.
This has an interesting generalisation to special holonomy spaces which arise as higher dimensional nilmanifolds fibred over a line \cite{Gibbons:2001ds}, giving
spaces of holonomy $SU(3),SU(4),G_2$ and $ Spin(7)$. 
In \cite{Chaemjumrus:2019wrt}, the solutions of 
 \cite{Gibbons:2001ds} were generalised  and, when combined with a Minkowski space factor, shown to be dual to intersecting brane solutions.
  In   \cite{Chaemjumrus:2019ipx,Chaemjumrus:2019wrt} it was also  seen that further,
dualities take these to T-folds \cite{Hull:2004in,Hull:2006va}
 and to locally non-geometric spaces with R-flux.
 
 The doubled formalism of \cite{Hull:2004in} provides a unified framework for treating all of these duals.
 The doubled space incorporates the original space and all of its T-duals. The various different duals arise by choosing different \lq polarisations' which select which half of the doubled coordinates are to be regarded as spacetime coordinates and which half are to be viewed as auxiliary winding coordinates. If the  solution does not depend on the winding coordinates in a given polarisation, the solution can be projected down to a spacetime solution, at least locally.
 If not, then, following \cite{Hull:2019iuy}, we refer to such solutions as {\it essentially doubled}.
 Spaces with R-flux are of this kind.
 T-duality can be viewed as changing the polarisation, and so changing which half of the doubled space is to be viewed as spacetime.
 
 String theory  in such doubled spaces was developed in \cite{Hull:2004in}-\cite{Siegel:1993xq} and the case of the doubling of group manifolds, twisted tori and nilmanifolds was first developed  
in \cite{Hull:2007jy,DallAgata:2007egc,Hull:2009sg,ReidEdwards:2009nu,Schulz:2011ye}. String theory can be formulated as a world-sheet theory of maps from a world-sheet into such a doubled space
 \cite{Hull:2004in}-\cite{Siegel:1993xq}.
  The choice of polarisation  and the consequent quotient can be obtained by a gauging of the doubled sigma model \cite{Hull:2006va}.
The doubled sigma model for group manifolds, twisted tori and nilmanifolds was developed  
in \cite{Hull:2007jy,Hull:2009sg,ReidEdwards:2009nu}
and the case of 
 the 3-dimensional nilfold was analysed in detail 
 in \cite{Hull:2009sg,ReidEdwards:2009nu}. For essentially doubled spaces, there is no way of projecting from the doubled formalism to a conventional string theory, and the doubled sigma model is then indispensible.
 
 Our aim here is to apply the framework of \cite{Hull:2009sg,ReidEdwards:2009nu} to the nilmanifolds 
 of  \cite{Gibbons:2001ds}  and to extend this to the string solutions that arise from fibring the nilmanifolds over a line.
 We double the nilmanifolds but leave the line undoubled (there are no winding modes on the line). In this way, we develop a formalism that works for all the duals, including the essentially doubled ones.
 
 Instead of working with the doubled sigma model, we could adopt a spacetime approach and use a formulation   in terms of double field theory 
 \cite{Hull:2009mi}-\cite{Hohm:2010pp}. The same doubled geometry will arise as a double field theory  configuration, and the choice of polarisation will correspond to a choice of solution of a non-linear version of the section constraint. The double field theory aspects will be discussed elsewhere.

\section{Scherk-Schwarz Reductions and the Doubled Formalism}

Dimensional reduction on a group $G$ has been well-studied. Using a Scherk-Schwarz ansatz  \cite{Scherk:1979zr}, the dependence of fields on the internal coordinates is given by a $G$ transformation and leads to a consistent truncation to a lower dimensional field theory; see \cite{Kaloper:1999yr,Hull:2005hk} for details of the Scherk-Schwarz dimensional reduction of the $\mathcal{N}$=1 supergravity action. However, for non-compact  $G$ this does not give a compactification, and considering  the full theory on such a  non-compact space gives a continuous spectrum without a mass gap.
In \cite{Hull:2005hk} it was argued that in order to lift the  Scherk-Schwarz dimensional reduction to a proper compactification of the full  string theory, it is necessary to instead consider 
compactification on a compact space $G/\Gamma$ with $\Gamma$ a discrete subgroup. A discrete  subgroup  that gives a compact quotient is said to be cocompact,
and so  this construction in string theory is restricted  to groups $G$ that admit a cocompact subgroup.

Lie groups $G$ that are nilpotent necessarily admit a cocompact subgroup and the resulting quotient  ${\cal N}= G/\Gamma$  is referred to as a nilmanifold.
A Lie group $G$   is nilpotent 
if  the  Lie algebra $\mathfrak{g}$   of $G$ satisfies
\begin{equation}
[X_1,[X_2,[\cdots[X_p,Y]\cdots]] = 0
\end{equation} 
for all $X_1, \cdots, X_p, Y \in \mathfrak{g}$, for some integer  $p $. For a nilpotent Lie group ${G}$, the smallest  such $p$ is known as the nilpotency class of ${G}$ and ${G}$ is then called a $p$-step nilpotent Lie group. A nilmanifold ${\cal N}= G/\Gamma$ 
is the  compact space given by the quotient of a nilpotent group $G$ by a cocompact discrete subgroup $\Gamma$.
For a $d$-dimensional  2-step nilpotent Lie group $G$ with centre of dimension $n$, the nilmanifold is a $T^n$ bundle over $T^m$ where $m=d-n$.
For example, the 3-dimensional Heisenberg group $G_3$ has 
a centre  of dimension $n=1$  and the quotient is known as the
nilfold and is an $S^1$ bundle over $T^2$.
Nilmanifolds are sometimes referred to as twisted tori.

The   Lie algebra generators $T_m$ 
satisfy an algebra
\begin{equation}
[T_m,T_n]=f_{mn}{}^p T_p
\end{equation}
and  the left-invariant one-forms
$P^m$ are
\begin{equation}
g^{-1}dg = P^mT_m.
\end{equation} 
Then 
the  general left-invariant metric on $ {G}$ is
\begin{equation}
ds^2 = x_{mn}P^mP^n,
\end{equation}
where $x_{mn}$ is a constant symmetric matrix, and descends to a metric on the nilmanifold ${\cal N}= G/\Gamma$ if $\Gamma$ is taken to act on the left.
Here $x_{mn}$ will be chosen as $x_{mn} = \delta_{mn}$ so  the metric is
\begin{equation}
ds^2 = \delta_{mn}P^mP^n.
\end{equation}

The  standard theory of gravity coupled to a 2-form gauge field  $B$ with field strength  $H=dB$ and dilaton $\Phi$ in $D$ dimensions has action
\begin{equation}
\label{D+d+1 lagrangian} S=\int \, e^{- 2{\Phi}}\left(  {R}*1-\frac{1}{2}
d {\Phi} \wedge *d {\Phi} -
\frac{1}{2}
{H}\wedge *H \right) \, .
\end{equation}
The Scherk-Schwarz reduction of this
for a group $G$ is given in \cite{Kaloper:1999yr,Hull:2005hk}.
In the abelian case $G=U(1)^d$, this gives a theory with $2d$ gauge fields, $d$ from the metric and $d$ from the $B$-field, 
and the gauge group is ${\cal G}=U(1)^{2d}$. There are $d^2$ scalar fields taking values in the coset $O(d,d)/O(d)\times O(d)$ in addition to the dilaton and the field theory in $D-d$ dimensions    has an $O(d,d)$ global symmetry.
For non-abelian $G$, the reduction gives a gauging of this theory, with 
gauge group 
$
G\ltimes \mathbb{R}^d$ 
with algebra
\begin{equation}
\label{dalg}
[T_m,T_n]=f_{mn}{}^p T_p, \qquad [T_m, \tilde T^n]= f_{mp}{}^n \tilde T^p, \qquad [\tilde T_m,\tilde T_n]=0
\end{equation}
where the factor $G$ generated by $T_m$ comes from the isometries of the group manifold generated by left-invariant vector fields and the abelian factor generated by
$\tilde{T}^m$ comes from the $B$-field symmetries.
This can be thought of as a gauging of a $2d$-dimensional subgroup of $O(d,d)$.
In addition, there is now a potential for  the $d^2+1$ scalar fields. Conditions for the potential  to have a Minkowski space critical point for an important class of cases were found in \cite{Dabholkar:2002sy};
  for generic groups  $G$,  Minkowski space in $D-d$ dimensions  will not be a  solution, but there will typically be domain wall solutions.

The Scherk-Schwarz reduction gives a truncation to a $D-d$ dimensional field theory.
This can be lifted to a compactification of the full supergravity or string theory
if $G$ admits a cocompact subgroup $\Gamma$.
Then the full string theory or supergravity theory can be compactified on $G/\Gamma$ and this
gives the same $D-d$ dimensional effective field theory but lifts to a compactification of the full supergravity or string theory.
 A 2-step nilmanifold  is a $T^n$ bundle over $T^m$, and this reduction can be regarded as compactification on $T^n$
followed by a reduction with duality twists on $T^m$, with a monodromy round each circle in $T^m$ that is a large diffeomorphism of $T^n$, in $SL(n,\mathbb{Z})$; this is then a compactification with duality twists \cite{Dabholkar:2002sy}.

For superstring theory, $D=10$ and (\ref{D+d+1 lagrangian}) is the action for   the massless graviton, dilaton and $B$-field of the type I, type II or heterotic  superstring. In this case, 
there is an interesting set of nilpotent groups $G$ such that the  resulting supergravity theory in $10-d$ dimensions has no Minkowski vacuum but has supersymmetric domain wall solutions \cite{Gibbons:2001ds}.
These each then lift to 10-dimensional solutions on ${\cal B}\times  \mathbb{R}^{1,r}$ where
$\mathbb{R}^{1,r}$ is $(r+1)$-dimensional Minkowski space with $r=8-d$ and ${\cal B}$ is  $G\times \mathbb{R}$ or 
${\cal N}\times \mathbb{R}$ \cite{Gibbons:2001ds}. Remarkably, as the domain wall was supersymmetric, the metric on ${\cal B}$ must have special holonomy, with the 
holonomy group determined by the number of supersymmetries \cite{Gibbons:2001ds}. For example for the Heisenberg group with ${\cal N}$ the nilfold, the four-dimensional  space 
${\cal B}$ 
is hyperk\" ahler. These cases were further analysed in \cite{Chaemjumrus:2019wrt} and we will focus on these examples in this paper. The   10-dimensional space   ${\cal B}\times  \mathbb{R}^{1,r}$ then incorporates the nilmanifold into a string solution.

In \cite{Chaemjumrus:2019wrt}, the T-duals of these solutions were considered. In each case,   T-dualities took the nilmanifold  to a torus $T^d$ with $H$-flux.
These T-dualities acting on the special holonomy domain wall solution ${\cal B}\times  \mathbb{R}^{1,r}$ resulted in a configuration of intersecting NS5-branes, preserving exactly the same amount of supersymmetry. Other T-dualities took these solutions to non-geometric backgrounds, including T-folds and spaces with {R}-flux.

These duals can thought of as follows.
If the dimension of the centre of $G$ is $n$, the nilmanifold is a $T^n$ bundle over $T^m$, but in each of the cases we will consider it can also be regarded as a $T^r$ bundle over $T^s$ for some $r>n$ with $s=d-r$, and we will take the maximal choice  of  $r$.
For example, the 3-dimensional nilfold can be regarded as a $T^2$ bundle over $S^1$.
The original nilmanifold compactification can  be regarded as compactification on $T^r$
followed by a reduction with duality twists on $T^s$, with a monodromy round each circle in $T^s$ that is  in $SL(r,\mathbb{Z})$.
There is an $O(r,r,\mathbb{Z})$ group of T-dualities acting on the $T^r$ fibres, and an $O(r,r,\mathbb{Z})$ transformation will take this to a
twisted reduction in which the monodromy round each circle in $T^s$ is now a   transformation in $O(r,r,\mathbb{Z})$.
On T-dualising to a torus $T^d$ with $H$-flux, the monodromies all consist of shifts of the $B$-field.
Other T-dualities   take it to cases in which the monodromies are T-dualities in $O(r,r,\mathbb{Z})$, giving a T-fold \cite{Hull:2004in}.

These duals can be represented in a  doubled formalism, in which the torus fibres $T^r$ are replaced with fibres that are given by a doubled torus $T^{2r}$, with an extra $r$ coordinates conjugate to the string winding modes on $T^r$.  The  $O(r,r,\mathbb{Z})$ monodromies  act geometrically as diffeomorphisms of the doubled torus  $T^{2r}$, so that a geometric $T^{2r}$ bundle over $T^s$ is obtained. This doubled solution can be thought of as a universal space containing all T-duals: different T-dual solutions are obtained by choosing different
polarisations, that is by choosing different splittings of the $2d$ coordinates into $d$ coordinates that are to be regarded  as the coordinates of a spacetime and  $d$ coordinates that are to be regarded as conjugate to winding numbers. T-duality can then be thought of as changing the polarisation \cite{Hull:2004in,Hull:2009sg}. This is worked out in detail for the nilfold in \cite{Hull:2009sg}.

T-duality on  the $T^s$ base is less straightforward and gives results that are not  geometric even locally and are sometimes said to have $R$-flux.
The metric of the nilmanifold ${\cal N}$ depends explicitly on the  coordinates $x^i$ of the $T^s$ base. T-duality takes the coordinate $x^i$ of the $i$'th circle  to the coordinate $\tilde x_i$ of the dual circle,  and so takes the original  $x$-dependent solution to one dependent on the dual coordinate $\tilde x_i$ \cite{Hull:2009sg,Dabholkar:2005ve}.
The monodromy round the original circle transforms to monodromy round the dual circle \cite{Hull:2009sg,Dabholkar:2005ve}. This was shown to give the correct T-duality in asymmetric orbifold limits in \cite{Dabholkar:2005ve}.
In such cases, the explicit  dependence of the solution on $\tilde x_i$  in general means there is no way of extracting a conventional background from the doubled one:
it is 
   essentially doubled.

In \cite{Hull:2009sg}, a doubled formulation of all these dualities was proposed. Instead of just doubling the torus fibres, all $d$ dimensions were doubled to give a space which is a $2d$ dimensional nilmanifold.
The Lie algebra (\ref{dalg}) is that of  $
{\cal G}=G\ltimes \mathbb{R}^d$, with group manifold given by the cotangent bundle $T^*G$ of $G$. This is itself a nilpotent group, and taking the quotient by a cocompact subgroup 
$\hat \Gamma$ gives a compact nilmanifold
${\cal M}= {\cal G}/ \hat \Gamma$.
This is  referred to as the doubled twisted torus.
The different choices of polarisation select the different dual backgrounds. This was checked in detail in \cite{Hull:2009sg} for the nilfold.
Our purpose here is to extend that to each of the nilmanifolds of \cite{Gibbons:2001ds}, constructing the doubled geometry on ${\cal M}= {\cal G}/ \hat \Gamma$, and extracting the various dual solutions. Then the special holonmy space  ${\cal B}={\cal N}\times \mathbb{R}$ is doubled to $\hat{\cal B}= {\cal M}\times \mathbb{R}$ -- there is no motivation to double the non-compact direction $\mathbb{R}$ as there is no winding mode and  no T-duality for this direction.


\section{The Doubled Nilmanifold} 

We now review the construction of  doubled nilmanifolds of \cite{Hull:2009sg}.
For  a nilmanifold $G/\Gamma$, the doubled group is 
the cotangent bundle ${\cal G}=T^*G$ of $G$, which is a Lie group
$
{\cal G}=G\ltimes \mathbb{R}^d$
with Lie algebra (\ref{dalg}), which we write as 
\begin{equation}
[T_M,T_N] = t_{MN}\,^PT_P. \label{Lie}
\end{equation}
where $M,N=1,\dots , 2d$.
The natural metric on the cotangent bundle is a metric $\eta_{MN}$ of signature $(d,d)$ invariant under ${\cal G}$
 and so ${\cal G}$ is  a subgroup of $O(d,d)$.

On $\mathcal{G}$, there are two sets of globally-defined  vector fields, the left-invariant vector fields, $K_M$, and the right-invariant vector fields, $\tilde{K}_M$. The left-invariant vector fields generate the right action $\mathcal{G}_R$, while the right-invariant vector fields generate left action $\mathcal{G}_L$. The left-invariant one-forms $\mathcal{P}^M$, dual to left-invariant vector field $K_M$, 
can be written as
\begin{equation}
g^{-1}dg = \mathcal{P}^M T_M
\end{equation} 
and
satisfy the Maurer-Cartan equations
\begin{equation}
d\mathcal{P}^M+\frac{1}{2}t_{NP}\,^M\mathcal{P}^N\wedge\mathcal{P}^P = 0.
\end{equation}

We introduce a left-invariant metric and three-form on $\mathcal{G}$   constructed from the left-invariant one-forms,  given by
\begin{eqnarray}\label{Gmetric}
ds^2 &=& \mathcal{M}_{MN}\mathcal{P}^M\otimes\mathcal{P}^N, \\
\mathcal{K} &=& \frac{1}{3!}t_{MNP}\mathcal{P}^M\wedge\mathcal{P}^N\wedge\mathcal{P}^P,\label{Kthreeform}
\end{eqnarray}
where $ \mathcal{M}_{MN}$ is a constant symmetric positive definite matrix and
$t_{MNP} = t_{MN}\,^Q\eta_{QP}$, and $t_{MNP} $ is totally antisymmetric.
The matrix $ \mathcal{M}_{MN}$  parameterizes the coset space $O(d,d)/O(d)\times O(d)$ and represents the  moduli of the internal space which become scalar fields in $10-d$ dimensions on compactification.

If $\Gamma '$ is a  cocompact subgroup of $\mathcal{G}$, 
taking the quotient gives a compact space
$\mathcal{M} = \mathcal{G}/\Gamma'$. If we take $\Gamma'$  to have a left action $g\to \gamma g$, then the left-invariant  1-forms $\mathcal{P}^M$, 
the metric (\ref{Gmetric}) and 3-form (\ref{Kthreeform})  descend to well-defined 1-forms, metric and 3-form on the quotient $\mathcal{M} $.
\subsection{The Sigma Model Formulation}

 In \cite{Hull:2009sg}, a doubled sigma model was formulated for maps from a 2-dimensional world-sheet $\Sigma$ to the doubled space $\mathcal{M} $.
 These maps pull back the one-forms $ \mathcal{P}^M$ 
 to one-forms $\hat {\mathcal{P}}^M$ on $\Sigma$.
 Introducing a 3-dimensional space  $S$ with boundary $\del S = \Sigma$ and extending the maps to $S$, the sigma model is 
    given by
\begin{equation}
S_{\mathcal{M}} = \frac{1}{4}\oint_\Sigma \mathcal{M}_{MN}
 \hat { \mathcal{P}}
^M \wedge * 
\hat {\mathcal{P}}
^N+\frac{1}{2}\int_S \hat { \mathcal{K}},
\end{equation}
where $ \hat { \mathcal{K}}$ is the pull-back of  $ \mathcal{K}$ to $S$ and $*$ is the Hodge dual on $\Sigma$.
This theory is subjected to the constraint
\begin{equation}
\hat { \mathcal{P}}^M = \eta ^{MP} \mathcal{M}_{PN} *\hat { \mathcal{P}}^N,
\end{equation}
which implies that half the degrees of freedom are right-moving on $\Sigma$ and half are left-moving.
This constraint can be imposed in a number of ways; in  \cite{Hull:2009sg} it was imposed by choosing a polarisation and then gauging to project to a quotient space.

\subsection{Polarisation and Section Condition}

A polarisation is a projector that projects the tangent space of $ \mathcal{G}$ into 
a physical subspace which is to be tangent to the 
subspace of the doubled space that is to be regarded as the physical 
spacetime. Different choices of polarisation select different dual backgrounds.
We introduce a projector $\Pi^m{}_M$ (with $m,n= 1,....,d$) mapping onto a $d$-dimensional subspace of the $2d$
dimensional tangent space of $ \mathcal{M}$,
which is totally null (maximally isotropic) with respect to the metric $\eta_{MN}$, i.e.
\begin{equation}
\eta^{MN} \Pi^m{}_M\Pi^n{}_N=0.
\end{equation}
Introducing such a projector at the identity element of the group manifold then defines one everywhere through the group action; in a natural frame, the projector is then constant over the manifold \cite{Hull:2004in}.
The complementary projector $1-\Pi$  is denoted by $ \widetilde{\Pi}_{mM}$.
The polarisation splits the tangent space into two halves, and we
will consider the case in which the frame components $ \Pi^m{}_M$ are locally constant, i.e. there is a constant matrix  $\Pi^m_{(\a)M}$ in each
patch $U_\a$ of $\cG$, but there can be different polarisation matrices in  different patches.

A vector $V^M$ is then projected into 
\begin{equation}
V^m= \Pi^m{}_M V^M, \qquad
V_m= \widetilde{\Pi}_{mM} V^M.
\end{equation}
It is useful to introduce the notation
\begin{equation}
V^{\hat M}=
 \left(%
\begin{array}{c}
V^m\\  V_m\\
\end{array}%
\right) 
= \Theta ^{\hat M}{}_N V^N ,
\end{equation}
where
\begin{equation}
\Theta ^{\hat M}{}_N = 
 \left(%
\begin{array}{c}
 \Pi^m{}_N\\  \widetilde{\Pi}_{mN}\\
\end{array}%
\right) ,
\end{equation}
so that the polarisation can be seen as choosing  a
basis for the tangent space.

The polarisation projects the   generators $T_M$
 into two sets, $Z_m$ and $X^m$:
\begin{equation}
\label{pgens}
Z_m \equiv \widetilde{\Pi}_{mM}\eta^{MN}T_N, \qquad X^m \equiv \Pi^m{}_M \eta ^{MN}T_N.
\end{equation}
The Lie algebra  (\ref{Lie})  then takes the form
\begin{equation}\nonumber
[Z_m,Z_n]=f_{mn}{}^pZ_p+K_{mnp}X^p, \qquad   [X^m,X^n]=Q_p{}^{mn}X^p+R^{mnp}Z_p,
\end{equation}
\begin{equation}
\label{palg}
[X^m,Z_n]=f_{np}{}^mX^p-Q_n{}^{mp}Z_p,
\end{equation}
for  some tensors $K_{mnp},f_{mn}{}^p,Q_p{}^{mn},R^{mnp}$, often referred to as fluxes, obtained by projecting the structure constants $t_{MN}{}^P$ with $\Pi, \widetilde{\Pi}$.
Different choices of polarisation will give different forms for these fluxes.

\subsection{Recovering the Physical Space}

For a given polarisation, we introduce coordinates $\mathbb{X}^M= (x^m, \tilde x_m)$
by writing a general group element 
as
\begin{equation}
g= \tilde h h,
\end{equation}
where 
\begin{equation}
h= \exp (x^mZ_m), \qquad \tilde h= \exp (\tilde x_mX^m).
\end{equation}
with $Z_m,X^m$ the projected generators (\ref{pgens}) defined with respect to the given polarisation.
The action of $h(x)$ on the generators $T_M = (Z_m, X^m)$ defines an $x$-dependent vielbein ${\cal V} _M {}^N(x)$ by
\begin{equation}
h^{-1} T_M h = {\cal V} _M {}^N T_N.
\end{equation}
Then defining
\begin{equation}
\Phi = \Phi ^M T_M =  \tilde h ^{-1} d  \tilde h + d h h^{-1}, \label{Phioneform}
\end{equation}
the left-invariant forms can be written as
\begin{equation}
 \mathcal{P}= \mathcal{P}^M T_M = \Phi ^M   {\cal V} _M {}^N(x)T_N. \label{PPhi}
\end{equation}
We define one-forms $ \tilde \ell ^m, \tilde \ell _m, r^m, r_m,\tilde{q}_ m$ by
\begin{equation}
 \tilde h ^{-1} d  \tilde h =  \tilde \ell ^mZ_m + \tilde \ell _m X^m, \qquad
  d h h^{-1} = r^mZ_m + r_m X^m,
\end{equation}
and
\begin{equation}
\tilde{q}_ m = r_m+\tilde{\ell}_m.
\end{equation}

 {A generalized metric which depends on the coordinates $x^i$ only} is defined by
\begin{equation}
{\cal H}_{MN}(x)={\cal M}_{PQ}{\cal V}^M{}_P{\cal V}^Q{}_N.\label{generalizemetric}
\end{equation}
Acting with  a polarisation tensor $\Theta_{\hat{M}}{}^M$
gives
\begin{equation}
{\cal H}_{\hat{M}\hat{N}}(x)=\Theta_{\hat{M}}{}^M{\cal H}_{MN}(x)\Theta^N{}_{\hat{N}},\label{hatGenmetric}
\end{equation}
 whose components define a metric $g_{mn}$ and $B$-field $B_{mn}$ by
\begin{equation}\label{H}
{\cal H}_{\hat{M}\hat{N}}(x)=
\left(\begin{array}{cc}
g_{mn}+B_{mp}g^{pq}B_{qn} & B_{mp}g^{pn} \\
g^{mp}B_{np} & g^{mn}
                          \end{array}\right).
\end{equation}
The metric $g_{mn}(x)$ and $B$-field $B_{mn}(x)$ depend only on the $x^i$ coordinates.
The physical metric is given by \cite{Hull:2009sg}
\begin{equation}\label{metricmn}
ds^2 = g_{mn}(x)r^mr^n.
\end{equation}
The physical $H$-field strength  is   given by \cite{Hull:2009sg}
\begin{equation}\label{H conjecture}
H=dB-\frac{1}{2}d\left(r^m\wedge {\tilde{q}}_m\right)+\frac{1}{2}{\cal K} \, .
\end{equation}

If the $R$-tensor $R^{mnp}$ vanishes, the $X^m$ generate a subgroup $\tilde G\subset \mathcal{G}$
\begin{equation}
[X^m,X^n]=Q_p{}^{mn}X^p.
\end{equation}
Then the generalized metric
${\cal H}_{MN}(x)$, the metric $g_{mn}$, the $B$-field $B_{mn}$ and the 3-form field strength $H$ are all invariant under the left action of $\widetilde{G}$.
The reduction to  the physical subspace is then  obtained by taking a quotient by
the left action of $\widetilde{G}$.
In the sigma model, this is achieved by gauging the action of $\widetilde{G}$. On eliminating the worldsheet gauge fields, one obtains a standard sigma model
whose  target space ${\mathcal G}/\widetilde{G}$ has coordinates $x$ and the  metric and $H$ given above.

If the $R$-tensor is not zero, then the model will depend explicitly on both $x$ and $\tilde x$. In this case, the expressions above give formal expressions for the metric and $H$ that depend on both $x$ and $\tilde x$, so that there is no interpretation in terms of a conventional $d$-dimensional spacetime. 

\subsection{Quotienting by the Discrete Group}

The above structure was derived  for the doubled group manifold ${\mathcal G}$.
In the case of a vanishing $R$-tensor, the result is a conventional sigma model on ${\mathcal G}/\widetilde{G}$.
The next step is to consider the structure for the nilmanifold $\mathcal{M}={\mathcal G}/\Gamma'$.

A conventional background is obtained from the doubled geometry by gauging $\tilde{X}^m$. The types of string theory background can be classified into three categories \cite{Hull:2009sg,ReidEdwards:2009nu}:

\noindent\underline{Type I:  Geometric Backgrounds}

\noindent If the   $\tilde {X}^m$ generate a subgroup $\tilde{G} \subset \mathcal{G}$ and this subgroup is preserved by $\Gamma'$, so that
\begin{equation}
\gamma k \gamma^{-1} = k',
\end{equation}
where $k, k' \in \tilde{G}$, $\gamma \in \Gamma'$, then the quotient space $\mathcal{M}/\tilde{G}$ is well-defined and gives a global description of a conventional geometric background.

\noindent\underline{Type II: T-fold Backgrounds}

\noindent If the   $\tilde {X}^m$ generate a subgroup $\tilde{G}$ but this subgroup is not preserved by $\Gamma'$, then the quotient space $\mathcal{M}/\tilde{G}$ is not well-defined. The conventional background can be recovered locally as a patch of $\mathcal{G}/\tilde{G}$. These patches are then   glued together together with T-duality transition functions, resulting in a T-fold.

\noindent\underline{Type III: Essentially doubled backgrounds}

\noindent
If the commutators of the $\tilde{X}^m$ do not close to generate a sub-algebra, then a conventional $d$-dimensional  background cannot be recovered even locally, as there is dependence on both $x^i$ and $\tilde x_i$, and the solution is essentially doubled. Such a background is   an {R}-flux background.

\section{The Nilfold Example}

The  doubled formalism was applied to    the 3-dimensional  nilfold  in \cite{Hull:2009sg}; here we summarize the results.
For this example, the nilpotent Lie group $G$ is  the three-dimensional Heisenberg group  with Lie algebra, generated by $T_i$ with $i=x,y,z$, given by
\begin{equation}
[T_x,T_z]=mT_y,  \qquad  [T_y,T_z]=0, \qquad [T_x,T_y]=0,
\end{equation}
with $m$ an integer.
The quotient of $G$ by a cocompact subgroup $\Gamma$ gives the nilfold ${\cal N}= G/\Gamma$.
Then the corresponding 6-dimensional group  $
{\cal G}=G\ltimes \mathbb{R}^3$, with generators $(T_i, \tilde T^i)$, 
has a Lie algebra whose only non-zero commutators are
\begin{equation}
[T_x,T_z]=mT_y,  \qquad   
\qquad  [T_x,\tilde T ^y]=m{\tilde{T}}^z,  \qquad  [T_z,\tilde T^y]=-m{\tilde{T}}^x. \label{nilfold1algebra}
\end{equation}
The cocompact subgroup $\Gamma '$ is given in \cite{Hull:2009sg} and consists of integer-valued matrices with respect to a suitable basis.

\subsection{The Nilfold}\label{NILFOLDsection}

Choosing the polarisation $\Theta = 1$, the algebra (\ref{nilfold1algebra}) written in terms of the projected generators (\ref{pgens}) is
\begin{equation}\label{nalgebra}
[Z_x,Z_z]=mZ_y,  \qquad  [Z_x,X^y]=mX^z,  \qquad  [Z_z,X^y]=-mX^x.
\end{equation}
The $X^m$ generate an abelian subgroup $\tilde G$, and taking the quotient by  $\tilde G$
gives the nilfold ${\cal N}= G/\Gamma$ with metric
\begin{equation}
ds^2_{\cal N}=dx^2+(dy-mxdz)^2+dz^2, 
\end{equation}
and $H=0$.
This can be viewed as a $T^2$ bundle over $S^1$ where the $T^2$ has coordinates $y,z$ and the $S^1$ has coordinate $x$.

\subsection{$T^3$ with $H$-flux}\label{HFLUXsection}

Choosing the polarisation
$$
\Theta=\left(%
\begin{array}{cccccc}
  1 & 0 & 0 & 0 & 0 & 0 \\
  0 & 0 & 0 & 0 & 1 & 0 \\
  0 & 0 & 1 & 0 & 0 & 0 \\
  0 & 0 & 0 & 1 & 0 & 0 \\
  0 & 1 & 0 & 0 & 0 & 0 \\
  0 & 0 & 0 & 0 & 0 & 1 \\
\end{array}%
\right),
$$
  the algebra written in terms of the projected generators (\ref{pgens}) is 
$$
[Z_x,Z_z]=mX^y,  \qquad  [Z_x,Z_y]=mX^z,  \qquad  [Z_z,Z_y]=-mX^x,
$$
with all other commutators vanishing.
The $X^i$ generate an abelian subgroup $\tilde G= \mathbb{R}^3$.
Taking the quotient gives
the 3-torus with $H$-flux given by the integer $m$. The metric and 3-form flux $H$ are
\begin{equation}
ds^2_{T^3}=dx^2+dy^2+dz^2,   \qquad  H=mdx\wedge dy\wedge dz.
\end{equation}
The 2-form potential $B$  with $H=dB$ 
can be chosen as
\begin{equation}
B=mxdy\wedge dz.
\end{equation}

\subsection{T-fold}\label{TFOLDsection}

The polarisation tensor for the T-fold is given by
$$
\Theta=\left(%
\begin{array}{cccccc}
  1 & 0 & 0 & 0 & 0 & 0 \\
  0 & 1 & 0 & 0 & 0 & 0 \\
  0 & 0 & 0 & 0 & 0 & 1 \\
  0 & 0 & 0 & 1 & 0 & 0 \\
  0 & 0 & 0 & 0 & 1 & 0 \\
  0 & 0 & 1 & 0 & 0 & 0 \\
\end{array}%
\right).
$$
The  algebra  written in terms of the projected generators (\ref{pgens}) is
then 
$$
[Z_x,X^z]=mZ_y,  \qquad  [Z_x,X^y]=mZ_z,  \qquad  [X^z,X^y]=-mX^x,
$$
where all other commutators vanish.
Now $X^x,X^y,X^z$ generate a  subgroup $\tilde G$ which is isomorphic to the Heisenberg group.
The T-fold
has
metric and $B$-field given by
\begin{equation}
ds^2_{T-Fold}=dx^2+\frac{1}{1+(mx)^2}(dy^2+dz^2),   \qquad  B=-\frac{mx}{1+(mx)^2}dy\wedge dz, 
\end{equation}
which changes by a T-duality under $x\to x+1$, and so
has a T-duality monodromy in the $x$ direction.

\subsection{Essentially Doubled Space}\label{RFLUXsection}

In this case, the  polarisation $\Theta$ is
$$
\Theta=\left(%
\begin{array}{cccccc}
  0 & 0 & 0 & 1 & 0 & 0 \\
  0 & 1 & 0 & 0 & 0 & 0 \\
  0 & 0 & 1 & 0 & 0 & 0 \\
  1 & 0 & 0 & 0 & 0 & 0 \\
  0 & 0 & 0 & 0 & 1 & 0 \\
  0 & 0 & 0 & 0 & 0 & 1 \\
\end{array}%
\right).
$$
The gauge algebra
is now
\begin{equation}
[{X}^x,{X}^z]=-m{Z}_y,  \qquad  [{X}^x,{X}^y]=-m{Z}_z,  \qquad
[{X}^z,{X}^y]=m{Z}_x,
\end{equation}
where all other commutators vanish.
Comparing with the general form of the Lie algebra (\ref{palg}), we see that
the structure constants $R^{mnp}$ are non-zero and so the space has $R$-flux.
The $X^i$ do not generate a subgroup, so there is no way of 
relating this  to a conventional theory on a $3$-dimensional space, even
locally, and the solution is essentially doubled.
\section{Higher Dimensional Nilmanifolds}\label{HigherNilfold} 

In this section, we construct the doubled geometry for each of the nilmanifolds considered in  \cite{Chaemjumrus:2019wrt}. Each nilmanifold is a $T^n$ bundle over $T^m$ for some $n,m$.

\subsection{$S^1$ bundle over $T^4$}
\label{S1T4}
For this example, the five-dimensional nilpotent Lie group has non-vanishing commutators
\begin{eqnarray}
\begin{array}{ll}
[T_2,T_3] = mT_1,& [T_4,T_5] = mT_1.
\end{array}
\end{eqnarray}
The corresponding ten-dimensional doubled group has a Lie algebra whose only non-zero commutators are
\begin{eqnarray}
\begin{array}{ll}
[T_2,T_3] = mT_1,& [T_4,T_5] = mT_1,\\
{[}\tilde{T}^1,T_2] = m\tilde{T}^3,& [\tilde{T}^1,T_3] = -m\tilde{T}^2,\\
{[}\tilde{T}^1,T_4] = m\tilde{T}^5,& [\tilde{T}^1,T_5] = -m\tilde{T}^4.\label{T1T4algebra}
\end{array}
\end{eqnarray}
The left-invariant one-forms are 
\begin{eqnarray}
\begin{array}{ll}
P^1 = dz^1+mz^3dz^2+mz^5dz^4,& Q_1 = d\tilde{z}_1,  \\
P^2 = dz^2, & Q_2 = d\tilde{z}_2-mz^3d\tilde{z}_1,\\
P^3 = dz^3,& Q_3 = d\tilde{z}_3+mz^2d\tilde{z}_1,\\
P^4 = dz^4, & Q_4 = d\tilde{z}_4-mz^5d\tilde{z}_1,\\
P^5 = dz^5, & Q_5 = d\tilde{z}_5+mz^4d\tilde{z}_1.
\end{array}
\end{eqnarray}

Choosing the polarisation $\Theta = 1$, the algebra (\ref{T1T4algebra}) in polarised form  is
\begin{eqnarray}
\begin{array}{ll}
[Z_2,Z_3] = mZ_1,& [Z_4,Z_5] = mZ_1,\\
{[}{X}^1,Z_2] = m{X}^3,& [{X}^1,Z_3] = -m{X}^2,\\
{[}{X}^1,Z_4] = m{X}^5,& [{X}^1,Z_5] = -m{X}^4.
\end{array}
\end{eqnarray}
The left-invariant one-forms in this polarization are 
\begin{eqnarray}
\begin{array}{ll}
P^1 = dz^1+mz^3dz^2+mz^5dz^4,& Q_1 = d\tilde{z}_1,  \\
P^2 = dz^2, & Q_2 = d\tilde{z}_2-mz^3d\tilde{z}_1,\\
P^3 = dz^3,& Q_3 = d\tilde{z}_3+mz^2d\tilde{z}_1,\\
P^4 = dz^4, & Q_4 = d\tilde{z}_4-mz^5d\tilde{z}_1,\\
P^5 = dz^5, & Q_5 = d\tilde{z}_5+mz^4d\tilde{z}_1.
\end{array}
\end{eqnarray}
Let $h = \prod\exp(z^mT_m)$ and $\tilde{h} = \prod\exp(\tilde{z}_m\tilde{T}^m)$, then the one-form (\ref{Phioneform}) is
\begin{equation}
\Phi = \Phi^M\,_NT^N = \tilde{h}^{-1}d\tilde{h}+dhh^{-1}.
\end{equation}
In this case, $dhh^{-1}$ and $\tilde{h}^{-1}d\tilde{h}$ are
\begin{equation}
dhh^{-1} = \Big(dz^1+mz^2dz^3+mz^4dz^5\Big)T_1+(dz^2)T_2+(dz^3)T_3+(dz^4)T_4+(dz^5)T_5,
\end{equation}
\begin{equation}
\tilde{h}^{-1}d\tilde{h} = (d\tilde{z}_1)\tilde{T}^1+(d\tilde{z}_2)\tilde{T}^2+(d\tilde{z}_3)\tilde{T}^3+(d\tilde{z}_4)\tilde{T}^4+(d\tilde{z}_5)\tilde{T}^5.
\end{equation}
The one-form $\Phi$ is given by 
\begin{eqnarray}
\begin{array}{ll}
\Phi^1 = dz^1+mz^2dz^3+mz^4dz^5,& \tilde{\Phi}_1 = d\tilde{z}_1,  \\
\Phi^2 = dz^2, & \tilde{\Phi}_2 = d\tilde{z}_2,\\
\Phi^3 = dz^3,& \tilde{\Phi}_3 = d\tilde{z}_3,\\
\Phi^4 = dz^4, & \tilde{\Phi}_4 = d\tilde{z}_4,\\
\Phi^5 = dz^5, & \tilde{\Phi}_5 = d\tilde{z}_5.
\end{array}
\end{eqnarray}
From the equation (\ref{PPhi}), one gets
\begin{equation}
\mathcal{V}^M\,_N=\left(%
\begin{array}{cccccccccc}
  1 & mz^3 & -mz^2 & mz^5 & -mz^4 & 0 & 0 & 0 & 0 & 0  \\
  0 & 1 & 0 & 0 & 0 & 0 & 0 & 0 & 0 & 0  \\
  0 & 0 & 1 & 0 & 0 & 0 & 0 & 0 & 0 & 0  \\
  0 & 0 & 0 & 1 & 0 & 0 & 0 & 0 & 0 & 0  \\
  0 & 0 & 0 & 0 & 1 & 0 & 0 & 0 & 0 & 0  \\
  0 & 0 & 0 & 0 & 0 & 1 & 0 & 0 & 0 & 0  \\
  0 & 0 & 0 & 0 & 0 & -mz^3 & 1 & 0 & 0 & 0  \\
  0 & 0 & 0 & 0 & 0 & mz^2 & 0 & 1 & 0 & 0  \\
  0 & 0 & 0 & 0 & 0 & -mz^5 & 0 & 0 & 1 & 0  \\
  0 & 0 & 0 & 0 & 0 & mz^4 & 0 & 0 & 0 & 1  \\
\end{array}%
\right),
\end{equation}
Using this $\mathcal{V}^M\,_N$, (\ref{generalizemetric}) gives the generalized metric $\mathcal{H}_{MN}$.
The $X^m$ generate an abelian subgroup $\tilde{G}$ and taking the quotient by $\tilde{G}$ gives the $S^1$ bundle over $T^4$ with metric
\begin{equation}
ds^2 = \Big(dz^1+m(z^3dz^2+z^5dz^4)\Big)^2+(dz^2)^2+(dz^3)^2+(dz^4)^2+(dz^5)^2. 
\end{equation}
The $H$-flux is given in (\ref{H conjecture}), which is
\begin{equation}
H=dB-\frac{1}{2}d\left(r^m\wedge {\tilde{q}}_m\right)+\frac{1}{2}{\cal K}.
\end{equation}
In this case, $\mathcal{K}$ and $r^m\wedge\tilde{q}_m$ are
\begin{equation}
\mathcal{K} = m\tilde{z}_1\wedge dz^2 \wedge dz^3+md\tilde{z}_1\wedge dz^4 \wedge dz^5,
\end{equation}
\begin{equation}
r^m\wedge\tilde{q}_m =\Big(dz^1+mz^2dz^3+mz^4dz^5\Big)\wedge d\tilde{z}_1+dz^2\wedge d\tilde{z}_2+dz^3\wedge d\tilde{z}_3+dz^4\wedge d\tilde{z}_4+dz^4\wedge d\tilde{z}_4,
\end{equation}
and $d(r^m\wedge\tilde{q}_m)$ is
\begin{equation}
d(r^m\wedge\tilde{q}_m) = m\tilde{z}_1\wedge dz^2 \wedge dz^3+md\tilde{z}_1\wedge dz^4 \wedge dz^5.
\end{equation}
In this case, $H$ = 0. A similar construction  works for the following choices of polarization.

 Choosing instead the polarization
\begin{equation}
\Theta=\left(%
\begin{array}{cccccccccc}
  0 & 0 & 0 & 0 & 0 & 1 & 0 & 0 & 0 & 0  \\
  0 & 1 & 0 & 0 & 0 & 0 & 0 & 0 & 0 & 0  \\
  0 & 0 & 1 & 0 & 0 & 0 & 0 & 0 & 0 & 0  \\
  0 & 0 & 0 & 1 & 0 & 0 & 0 & 0 & 0 & 0  \\
  0 & 0 & 0 & 0 & 1 & 0 & 0 & 0 & 0 & 0  \\
  1 & 0 & 0 & 0 & 0 & 0 & 0 & 0 & 0 & 0  \\
  0 & 0 & 0 & 0 & 0 & 0 & 1 & 0 & 0 & 0  \\
  0 & 0 & 0 & 0 & 0 & 0 & 0 & 1 & 0 & 0  \\
  0 & 0 & 0 & 0 & 0 & 0 & 0 & 0 & 1 & 0  \\
  0 & 0 & 0 & 0 & 0 & 0 & 0 & 0 & 0 & 1  \\
\end{array}%
\right),
\end{equation}
the algebra (\ref{T1T4algebra}) is written in terms of the projected generators as
\begin{eqnarray}
\begin{array}{ll}
[Z_2,Z_3] = mX^1,& [Z_4,Z_5] = mX^1,\\
{[}Z_1,Z_2] = m{X}^3,& [Z_1,Z_3] = -m{X}^2,\\
{[}Z_1,Z_4] = m{X}^5,& [Z_1,Z_5] = -m{X}^4.
\end{array}
\end{eqnarray}
The $X^m$ generate an abelian subgroup $\tilde{G}$. Taking the quotient by $\tilde G$ gives the $5$-torus with $H$-flux
\begin{eqnarray}
ds^2 &=& (dz^1)^2+(dz^2)^2+(dz^3)^2+(dz^4)^2+(dz^5)^2,\\
H &=& -mdz^1\wedge dz^2 \wedge dz^3 - mdz^1 \wedge dz^4 \wedge dz^5ใ
\end{eqnarray}

 Choosing the polarization
\begin{equation}
\Theta=\left(%
\begin{array}{cccccccccc}
  1 & 0 & 0 & 0 & 0 & 0 & 0 & 0 & 0 & 0  \\
  0 & 0 & 0 & 0 & 0 & 0 & 1 & 0 & 0 & 0  \\
  0 & 0 & 1 & 0 & 0 & 0 & 0 & 0 & 0 & 0  \\
  0 & 0 & 0 & 0 & 0 & 0 & 0 & 0 & 1 & 0  \\
  0 & 0 & 0 & 0 & 1 & 0 & 0 & 0 & 0 & 0  \\
  0 & 0 & 0 & 0 & 0 & 1 & 0 & 0 & 0 & 0  \\
  0 & 1 & 0 & 0 & 0 & 0 & 0 & 0 & 0 & 0  \\
  0 & 0 & 0 & 0 & 0 & 0 & 0 & 1 & 0 & 0  \\
  0 & 0 & 0 & 1 & 0 & 0 & 0 & 0 & 0 & 0  \\
  0 & 0 & 0 & 0 & 0 & 0 & 0 & 0 & 0 & 1  \\
\end{array}%
\right),
\end{equation}
the algebra (\ref{T1T4algebra}) is
\begin{eqnarray}
\begin{array}{ll}
[X^2,Z_3] = mZ_1,& [X^4,Z_5] = mZ_1,\\
{[}X^1,X^2] = m{X}^3,& [X^1,Z_3] = -mZ_2,\\
{[}X^1,X^4] = m{X}^5,& [X^1,Z_5] = -mZ_4.
\end{array}
\end{eqnarray}
The $X^m$ generate a subgroup $\tilde{G}$. Taking the quotient gives the T-fold with the metric and $B$-field
\begin{eqnarray}
ds^2 &=& \frac{1}{1+m^2\Big[(z^3)^2+(z^5)^2\Big]}\Big((dz^1)^2+(dz^2)^2+(dz^4)^2\Big)\nonumber\\
& &+\frac{1}{1+m^2\Big[(z^3)^2+(z^5)^2\Big]}\Big(mz^5dz^2-mz^3dz^4\Big)^2+(dz^3)^2+(dz^5)^2,\\
B &=& \frac{m}{1+m^2\Big[(z^3)^2+(z^5)^2\Big]}\Big(z^3dz^1\wedge dz^2+z^5 dz^1\wedge dz^4\Big).
\end{eqnarray}

\subsection{$T^2$ bundle over $T^3$}
\label{T2T3}
For this example, the five-dimensional nilpotent Lie group has non-vanishing commutators
\begin{equation}
\begin{array}{ll}
[T_3,T_4] = mT_1,& [T_3,T_5] = mT_2.
\end{array}
\end{equation}
The corresponding ten-dimensional group has a Lie algebra whose only non-zero commutators are
\begin{eqnarray}
\begin{array}{ll}
[T_3,T_4] = mT_1,& [T_3,T_5] = mT_2,\\
{[}\tilde{T}^1,T_3] = m\tilde{T}^4,& [\tilde{T}^1,T_4] = -m\tilde{T}^3,\\
{[}\tilde{T}^2,T_3] = m\tilde{T}^5,& [\tilde{T}^2,T_5] = -m\tilde{T}^3.
\end{array}\label{T2T3algebra}
\end{eqnarray}
The left-invariant one-forms are 
\begin{eqnarray}
\begin{array}{ll}
P^1 = dz^1+mz^4dz^3,& Q_1 = d\tilde{z}_1,  \\
P^2 = dz^2+mz^5dz^3, & Q_2 = d\tilde{z}_2,\\
P^3 = dz^3, & Q_3 = d\tilde{z}_3+mz^4d\tilde{z}_1-mz^5d\tilde{z}_2,\\
P^4 = dz^4, & Q_4 = d\tilde{z}_4+mz^3d\tilde{z}_1,\\
P^5 = dz^5, & Q_5 = d\tilde{z}_5+mz^3d\tilde{z}_2.
\end{array}
\end{eqnarray}

Choosing the polarisation $\Theta = 1$, the algebra (\ref{T2T3algebra}) is
\begin{eqnarray}
\begin{array}{ll}
[Z_3,Z_4] = mZ_1,& [Z_3,Z_5] = mZ_2,\\
{[}{X}^1,Z_3] = m{X}^4,& [{X}^1,Z_4] = -m{X}^3,\\
{[}{X}^2,Z_3] = m{X}^5,& [{X}^2,Z_5] = -m{X}^3.
\end{array}
\end{eqnarray}
The $X^m$ generate an abelian subgroup $\tilde{G}$ and taking the quotient by $\tilde{G}$ gives the $T^2$ bundle over $T^3$ with metric
\begin{equation}
ds^2 = \Big(dz^1+mz^4dz^3\Big)^2+\Big(dz^2+mz^5dz^3\Big)^2+(dz^3)^2+(dz^4)^2+(dz^5)^2. 
\end{equation}

Choosing the polarization
\begin{equation}
\Theta=\left(%
\begin{array}{cccccccccc}
  0 & 0 & 0 & 0 & 0 & 1 & 0 & 0 & 0 & 0  \\
  0 & 1 & 0 & 0 & 0 & 0 & 0 & 0 & 0 & 0  \\
  0 & 0 & 1 & 0 & 0 & 0 & 0 & 0 & 0 & 0  \\
  0 & 0 & 0 & 1 & 0 & 0 & 0 & 0 & 0 & 0  \\
  0 & 0 & 0 & 0 & 1 & 0 & 0 & 0 & 0 & 0  \\
  1 & 0 & 0 & 0 & 0 & 0 & 0 & 0 & 0 & 0  \\
  0 & 0 & 0 & 0 & 0 & 0 & 1 & 0 & 0 & 0  \\
  0 & 0 & 0 & 0 & 0 & 0 & 0 & 1 & 0 & 0  \\
  0 & 0 & 0 & 0 & 0 & 0 & 0 & 0 & 1 & 0  \\
  0 & 0 & 0 & 0 & 0 & 0 & 0 & 0 & 0 & 1  \\
\end{array}%
\right),
\end{equation}
the algebra (\ref{T2T3algebra}) is
\begin{eqnarray}
\begin{array}{ll}
[Z_3,Z_4] = mX^1,& [Z_3,Z_5] = mZ_2,\\
{[}Z_1,Z_3] = m{X}^4,& [Z_1,Z_4] = -m{X}^3,\\
{[}{X}^2,Z_3] = m{X}^5,& [{X}^2,Z_5] = -m{X}^3.
\end{array}
\end{eqnarray}
The $X^m$ generate an abelian subgroup $\tilde{G}$. Taking the quotient gives the $S^1$ bundle over $T^4$ with $H$-flux.
\begin{eqnarray}
ds^2 &=& (dz^1)^2+\Big(dz^2+mz^5dz^3\Big)^2+(dz^3)^2+(dz^4)^2+(dz^5)^2,\\
H &=& - mdz^1 \wedge dz^3 \wedge dz^4.
\end{eqnarray}

For the polarization
\begin{equation}
\Theta=\left(%
\begin{array}{cccccccccc}
  0 & 0 & 0 & 0 & 0 & 1 & 0 & 0 & 0 & 0  \\
  0 & 0 & 0 & 0 & 0 & 0 & 1 & 0 & 0 & 0  \\
  0 & 0 & 1 & 0 & 0 & 0 & 0 & 0 & 0 & 0  \\
  0 & 0 & 0 & 1 & 0 & 0 & 0 & 0 & 0 & 0  \\
  0 & 0 & 0 & 0 & 1 & 0 & 0 & 0 & 0 & 0  \\
  1 & 0 & 0 & 0 & 0 & 0 & 0 & 0 & 0 & 0  \\
  0 & 1 & 0 & 0 & 0 & 0 & 0 & 0 & 0 & 0  \\
  0 & 0 & 0 & 0 & 0 & 0 & 0 & 1 & 0 & 0  \\
  0 & 0 & 0 & 0 & 0 & 0 & 0 & 0 & 1 & 0  \\
  0 & 0 & 0 & 0 & 0 & 0 & 0 & 0 & 0 & 1  \\
\end{array}%
\right),
\end{equation}
the algebra (\ref{T2T3algebra}) is
\begin{eqnarray}
\begin{array}{ll}
[Z_3,Z_4] = mX^1,& [Z_3,Z_5] = mX^2,\\
{[}Z_1,Z_3] = m{X}^4,& [Z_1,Z_4] = -m{X}^3,\\
{[}Z_2,Z_3] = m{X}^5,& [Z_2,Z_5] = -m{X}^3.
\end{array}
\end{eqnarray}
The $X^m$ generate an abelian subgroup $\tilde{G}$. Taking the quotient gives the $T^5$ with $H$-flux.
\begin{eqnarray}
ds^2 &=& (dz^1)^2+(dz^2)^2+(dz^3)^3+(dz^4)^2+(dz^5)^2,\\
H &=& -mdz^1\wedge dz^3\wedge dz^4-mdz^2\wedge dz^3 \wedge dz^5.
\end{eqnarray}

Choosing the polarization
\begin{equation}
\Theta=\left(%
\begin{array}{cccccccccc}
  1 & 0 & 0 & 0 & 0 & 0 & 0 & 0 & 0 & 0  \\
  0 & 1 & 0 & 0 & 0 & 0 & 0 & 0 & 0 & 0  \\
  0 & 0 & 0 & 0 & 0 & 0 & 0 & 1 & 0 & 0  \\
  0 & 0 & 0 & 1 & 0 & 0 & 0 & 0 & 0 & 0  \\
  0 & 0 & 0 & 0 & 1 & 0 & 0 & 0 & 0 & 0  \\
  0 & 0 & 0 & 0 & 0 & 1 & 0 & 0 & 0 & 0  \\
  0 & 0 & 0 & 0 & 0 & 0 & 1 & 0 & 0 & 0  \\
  0 & 0 & 1 & 0 & 0 & 0 & 0 & 0 & 0 & 0  \\
  0 & 0 & 0 & 0 & 0 & 0 & 0 & 0 & 1 & 0  \\
  0 & 0 & 0 & 0 & 0 & 0 & 0 & 0 & 0 & 1  \\
\end{array}%
\right),
\end{equation}
the algebra (\ref{T2T3algebra}) is
\begin{eqnarray}
\begin{array}{ll}
[X^3,Z_4] = mZ_1,& [X^3,Z_5] = mZ_2,\\
{[}{X}^1,X^3] = m{X}^4,& [{X}^1,Z_4] = -mZ_3,\\
{[}{X}^2,X^3] = m{X}^5,& [{X}^2,Z_5] = -mZ_3.
\end{array}
\end{eqnarray}
The $X^m$ generate a subgroup $\tilde{G}$. Taking the quotient gives the T-fold with the metric and $B$-field
\begin{eqnarray}
ds^2 &=& \frac{1}{1+m^2\Big[(z^4)^2+(z^5)^2\Big]}\Big[(dz^1)^2+(dz^2)^2+(dz^3)^2\Big]+\frac{1}{1+m^2\Big[(z^4)^2+(z^5)^2\Big]}\Big(z^5dz^1-z^4dz^2\Big)^2\nonumber\\
& & +(dz^4)^2+(dz^5)^2,\\
B &=& \frac{m}{1+m^2\Big[(z^4)^2+(z^5)^2\Big]}\Big(z^4dz^1\wedge dz^3+z^5dz^2\wedge dz^3\Big).
\end{eqnarray}

\subsection{$T^2$ bundle over $T^4$}
\label{T2T4}
Consider the six-dimensional nilpotent Lie algebra whose  only non-vanishing commutators are
\begin{eqnarray}
\begin{array}{ll}
[T_3,T_4] = mT_1,& [T_3,T_5] = mT_2,\\
{[T_5,T_6]} = mT_1,& [T_4,T_6] = -mT_2.
\end{array}
\end{eqnarray}
The corresponding twelve-dimensional group has a Lie algebra whose only non-zero commutators are
\begin{eqnarray}
\begin{array}{ll}
[T_3,T_4] = mT_1,&[T_3,T_5] = mT_2,\\
{[T_5,T_6]} = mT_1,& [T_4,T_6] = -mT_2,\\
{[}\tilde{T}^1,T_3] = m\tilde{T}^4,& [\tilde{T}^1,T_4] = -m\tilde{T}^3,\\
{[}\tilde{T}^1,T_5] = m\tilde{T}^6,& [\tilde{T}^1,T_6] = -m\tilde{T}^5,\\
{[}\tilde{T}^2,T_3] = m\tilde{T}^5,& [\tilde{T}^2,T_5] = -m\tilde{T}^3,\\
{[}\tilde{T}^2,T_6] = m\tilde{T}^4,& [\tilde{T}^2,T_4] = -m\tilde{T}^6.\label{T2T4algebra}
\end{array}
\end{eqnarray}
The left-invariant one-forms are 
\begin{eqnarray}
\begin{array}{ll}
P^1 = dz^1+mz^4dz^3+mz^6dz^5,& Q_1 = d\tilde{z}_1,  \\
P^2 = dz^2+mz^5dz^3-mz^6dz^4, & Q_2 = d\tilde{z}_2,\\
P^3 = dz^3, & Q_3 = d\tilde{z}_3-mz^4d\tilde{z}_1-mz^5d\tilde{z}_2,\\
P^4 = dz^4, & Q_4 = d\tilde{z}_4+mz^3d\tilde{z}_1+mz^6d\tilde{z}_2,\\
P^5 = dz^5, & Q_5 = d\tilde{z}_5-mz^6d\tilde{z}_1+mz^3d\tilde{z}_2,\\
P^6 = dz^6, & Q_6 = d\tilde{z}_6+mz^6d\tilde{z}_1-mz^4d\tilde{z}_2.
\end{array}
\end{eqnarray}

Choosing the polarisation $\Theta = 1$, the algebra (\ref{T2T4algebra}) is
\begin{eqnarray}
\begin{array}{ll}
[Z_3,Z_4] = mZ_1,&[Z_3,Z_5] = mZ_2,\\
{[Z_5,Z_6]} = mZ_1,& [Z_4,Z_6] = -mZ_2,\\
{[}{X}^1,Z_3] = m{X}^4,& [{X}^1,Z_4] = -m{X}^3,\\
{[}{X}^1,Z_5] = m{X}^6,& [{X}^1,Z_6] = -m{X}^5,\\
{[}{X}^2,Z_3] = m{X}^5,& [{X}^2,Z_5] = -m{X}^3,\\
{[}{X}^2,Z_6] = m{X}^4,& [{X}^2,Z_4] = -m{X}^6.
\end{array}
\end{eqnarray}
The $X^m$ generate an abelian subgroup $\tilde{G}$. Taking the quotient by $\tilde{G}$ gives the metric
\begin{equation}
ds^2 = \Big(dz^1+m(z^4dz^3+z^6dz^5)\Big)^2+\Big(dz^2+m(z^5dz^3-z^6dz^4)\Big)^2+(dz^3)^2+(dz^4)^2+(dz^5)^2+(dz^6)^2. 
\end{equation}

Choosing the polarization
\begin{equation}
\Theta=\left(%
\begin{array}{cccccccccccc}
  0 & 0 & 0 & 0 & 0 & 0 & 1 & 0 & 0 & 0 & 0 & 0 \\
 0 & 0 & 0 & 0 & 0 & 0 & 0 & 1 & 0 & 0 & 0 & 0 \\
 0 & 0 & 1 & 0 & 0 & 0 & 0 & 0 & 0 & 0 & 0 & 0 \\
 0 & 0 & 0 & 1 & 0 & 0 & 0 & 0 & 0 & 0 & 0 & 0 \\
  0 & 0 & 0 & 0 & 1 & 0 & 0 & 0 & 0 & 0 & 0 & 0 \\
 0 & 0 & 0 & 0 & 0 & 1 & 0 & 0 & 0 & 0 & 0 & 0 \\
 1 & 0 & 0 & 0 & 0 & 0 & 0 & 0 & 0 & 0 & 0 & 0 \\
 0 & 1 & 0 & 0 & 0 & 0 & 0 & 0 &0 & 0 & 0 & 0 \\
 0 & 0 & 0 & 0 & 0 & 0 & 0 & 0 & 1 & 0 & 0 & 0 \\
 0 & 0 & 0 & 0 & 0 & 0 & 0 & 0 & 0 & 1 & 0 & 0 \\
 0 & 0 & 0 & 0 & 0 & 0 & 0 & 0 & 0 & 0 & 1 & 0 \\
  0 & 0 & 0 & 0 & 0 & 0 & 0 & 0 & 0 & 0 & 0 & 1 \\
\end{array}%
\right),
\end{equation}
the algebra (\ref{T2T4algebra}) is
\begin{eqnarray}
\begin{array}{ll}
[Z_3,Z_4] = mX^1,&[Z_3,Z_5] = mX^2,\\
{[Z_5,Z_6]} = mX^1,& [Z_4,Z_6] = -mX^2,\\
{[}Z_1,Z_3] = m{X}^4,& [Z_1,Z_4] = -m{X}^3,\\
{[}Z_1,Z_5] = m{X}^6,& [Z_1,Z_6] = -m{X}^5,\\
{[}Z_2,Z_3] = m{X}^5,& [Z_2,Z_5] = -m{X}^3,\\
{[}Z_2,Z_6] = m{X}^4,& [Z_2,Z_4] = -m{X}^6.
\end{array}
\end{eqnarray}
The $X^m$ generate an abelian subgroup $\tilde{G}$. Taking the quotient gives the $T^6$ with $H$-flux.
\begin{eqnarray}
ds^2 = (dz^1)^2+(dz^2)^2+(dz^3)^2+(dz^4)^2+(dz^5)^2+(dz^6)^2,
\end{eqnarray}
and $H$-flux
\begin{eqnarray}
H &=&-mdz^1 \wedge dz^3 \wedge dz^4 -mdz^1 \wedge dz^5 \wedge dz^6\nonumber\\
& & -mdz^2 \wedge dz^3 \wedge dz^5 +mdz^2\wedge dz^4 \wedge dz^6.
\end{eqnarray}

For the polarization
\begin{equation}
\Theta=\left(%
\begin{array}{cccccccccccc}
  1 & 0 & 0 & 0 & 0 & 0 & 0 & 0 & 0 & 0 & 0 & 0 \\
 0 & 1 & 0 & 0 & 0 & 0 & 0 & 0 & 0 & 0 & 0 & 0 \\
 0 & 0 & 0 & 0 & 0 & 0 & 0 & 0 & 1 & 0 & 0 & 0 \\
 0 & 0 & 0 & 1 & 0 & 0 & 0 & 0 & 0 & 0 & 0 & 0 \\
  0 & 0 & 0 & 0 & 1 & 0 & 0 & 0 & 0 & 0 & 0 & 0 \\
 0 & 0 & 0 & 0 & 0 & 1 & 0 & 0 & 0 & 0 & 0 & 0 \\
 0 & 0 & 0 & 0 & 0 & 0 & 1 & 0 & 0 & 0 & 0 & 0 \\
 0 & 0 & 0 & 0 & 0 & 0 & 0 & 1 &0 & 0 & 0 & 0 \\
 0 & 0 & 1 & 0 & 0 & 0 & 0 & 0 & 0 & 0 & 0 & 0 \\
 0 & 0 & 0 & 0 & 0 & 0 & 0 & 0 & 0 & 1 & 0 & 0 \\
 0 & 0 & 0 & 0 & 0 & 0 & 0 & 0 & 0 & 0 & 1 & 0 \\
  0 & 0 & 0 & 0 & 0 & 0 & 0 & 0 & 0 & 0 & 0 & 1 \\
\end{array}%
\right),
\end{equation}
the algebra (\ref{T2T4algebra}) is
\begin{eqnarray}
\begin{array}{ll}
[X^3,Z_4] = mZ_1,&[X^3,Z_5] = mZ_2,\\
{[Z_5,Z_6]} = mZ_1,& [Z_4,Z_6] = -mZ_2,\\
{[}{X}^1,X^3] = m{X}^4,& [{X}^1,Z_4] = -mZ_3,\\
{[}{X}^1,Z_5] = m{X}^6,& [{X}^1,Z_6] = -m{X}^5,\\
{[}{X}^2,X^3] = m{X}^5,& [{X}^2,Z_5] = -mZ_3,\\
{[}{X}^2,Z_6] = m{X}^4,& [{X}^2,Z_4] = -m{X}^6.
\end{array}
\end{eqnarray}
The $X^m$ generate a subgroup $\tilde{G}$. Taking the quotient gives
a T-fold with metric and $B$-field
given by (\ref{T-fold1metric}) and (\ref{T-fold1Bfield}).

\subsection{$T^3$ bundle over $T^3$}
\label{T3T3}
Consider the six-dimensional nilpotent Lie algebra whose  only non-vanishing commutators are
\begin{eqnarray}
\begin{array}{ll}
[T_5,T_6] = mT_1,& [T_4,T_6] = -mT_2,\\
{[T_4,T_5]} = mT_3.&
\end{array}
\end{eqnarray}
The corresponding twelve-dimensional group has a Lie algebra whose only non-zero commutators are
\begin{eqnarray}
\begin{array}{ll}
[T_5,T_6] = mT_1,& [T_4,T_6] = -mT_2,\\
{[T_4,T_5]} = mT_3,&[\tilde{T}^1,T_5] = m\tilde{T}^6,\\
{[}\tilde{T}^2,T_6] = m\tilde{T}^4,& [\tilde{T}^2,T_4] = -m\tilde{T}^6,\\
{[}\tilde{T}^3,T_4] = m\tilde{T}^5,& [\tilde{T}^1,T_6] = -m\tilde{T}^5,\\
{[}\tilde{T}^3,T_5] = -m\tilde{T}^4.& \label{T3T3algebra}
\end{array}
\end{eqnarray}
The left-invariant one-forms are 
\begin{eqnarray}
\begin{array}{ll}
P^1 = dz^1+mz^6dz^5,& Q_1 = d\tilde{z}_1,  \\
P^2 = dz^2-mz^6dz^4, & Q_2 = d\tilde{z}_2,\\
P^3 = dz^3+mz^5dz^4, & Q_3 = d\tilde{z}_3,\\
P^4 = dz^4, & Q_4 = d\tilde{z}_4+mz^6d\tilde{z}_2+mz^5d\tilde{z}_3,\\
P^5 = dz^5, & Q_5 = d\tilde{z}_5-mz^6d\tilde{z}_1+mz^4d\tilde{z}_3,\\
P^6 = dz^6, & Q_6 = d\tilde{z}_6+mz^5d\tilde{z}_1-mz^4d\tilde{z}_2.
\end{array}
\end{eqnarray}

Choosing the polarisation $\Theta = 1$, the algebra (\ref{T3T3algebra}) is
\begin{eqnarray}
\begin{array}{ll}
[Z_5,Z_6] = mZ_1,& [Z_4,Z_6] = -mZ_2,\\
{[Z_4,Z_5]} = mZ_3,&[{X}^1,Z_5] = m{X}^6,\\
{[}{X}^2,Z_6] = m{X}^4,& [{X}^2,Z_4] = -m{X}^6,\\
{[}{X}^3,Z_4] = m{X}^5,& [{X}^1,Z_6] = -m{X}^5,\\
{[}{X}^3,Z_5] = -m{X}^4.& 
\end{array}
\end{eqnarray}
The $X^m$ generate an abelian subgroup $\tilde{G}$. Taking the quotient by $\tilde{G}$ gives the metric
\begin{equation}
ds^2 = \Big(dz^1+mz^6dz^5\Big)^2+\Big(dz^2-m^2z^6dz^4\Big)^2+\Big(dz^3+mz^5dz^4\Big)^2+(dz^4)^2+(dz^5)^2+(dz^6)^2.
\end{equation}

Choosing the polarization
\begin{equation}
\Theta=\left(%
\begin{array}{cccccccccccc}
  0 & 0 & 0 & 0 & 0 & 0 & 1 & 0 & 0 & 0 & 0 & 0 \\
 0 & 0 & 0 & 0 & 0 & 0 & 0 & 1 & 0 & 0 & 0 & 0 \\
 0 & 0 & 0 & 0 & 0 & 0 & 0 & 0 & 1 & 0 & 0 & 0 \\
 0 & 0 & 0 & 1 & 0 & 0 & 0 & 0 & 0 & 0 & 0 & 0 \\
  0 & 0 & 0 & 0 & 1 & 0 & 0 & 0 & 0 & 0 & 0 & 0 \\
 0 & 0 & 0 & 0 & 0 & 1 & 0 & 0 & 0 & 0 & 0 & 0 \\
 1 & 0 & 0 & 0 & 0 & 0 & 0 & 0 & 0 & 0 & 0 & 0 \\
 0 & 1 & 0 & 0 & 0 & 0 & 0 & 0 &0 & 0 & 0 & 0 \\
 0 & 0 & 1 & 0 & 0 & 0 & 0 & 0 & 0 & 0 & 0 & 0 \\
 0 & 0 & 0 & 0 & 0 & 0 & 0 & 0 & 0 & 1 & 0 & 0 \\
 0 & 0 & 0 & 0 & 0 & 0 & 0 & 0 & 0 & 0 & 1 & 0 \\
  0 & 0 & 0 & 0 & 0 & 0 & 0 & 0 & 0 & 0 & 0 & 1 \\
\end{array}%
\right),
\end{equation}
the algebra (\ref{T3T3algebra}) is
\begin{eqnarray}
\begin{array}{ll}
[Z_5,Z_6] = mX^1,& [Z_4,Z_6] = -mX^2,\\
{[Z_4,Z_5]} = mX^3,&[Z_1,Z_5] = m{X}^6,\\
{[}{Z}_2,Z_6] = m{X}^4,& [{Z}_2,Z_4] = -m{X}^6,\\
{[}{Z}_3,Z_4] = m{X}^5,& [{Z}_1,Z_6] = -m{X}^5,\\
{[}{Z}_3,Z_5] = -m{X}^4.& 
\end{array}
\end{eqnarray}
The $X^m$ generate an abelian subgroup $\tilde{G}$. Taking the quotient by $\tilde{G}$ gives a $T^6$  with $H$-flux
\begin{eqnarray}
ds^2 = (dz^1)^2+(dz^2)^2+(dz^3)^2+(dz^4)^2+(dz^5)^2+(dz^6)^2,
\end{eqnarray}
\begin{equation}
H = -m dz^1\wedge dz^5 \wedge dz^6 - mdz^3 \wedge dz^4\wedge dz^5 +mdz^2\wedge dz^4 \wedge dz^6.
\end{equation}

Choosing the polarization
\begin{equation}
\Theta=\left(%
\begin{array}{cccccccccccc}
  1 & 0 & 0 & 0 & 0 & 0 & 0 & 0 & 0 & 0 & 0 & 0 \\
 0 & 1 & 0 & 0 & 0 & 0 & 0 & 0 & 0 & 0 & 0 & 0 \\
 0 & 0 & 1 & 0 & 0 & 0 & 0 & 0 & 0 & 0 & 0 & 0 \\
 0 & 0 & 0 & 0 & 0 & 0 & 0 & 0 & 0 & 1 & 0 & 0 \\
  0 & 0 & 0 & 0 & 1 & 0 & 0 & 0 & 0 & 0 & 0 & 0 \\
 0 & 0 & 0 & 0 & 0 & 1 & 0 & 0 & 0 & 0 & 0 & 0 \\
 0 & 0 & 0 & 0 & 0 & 0 & 1 & 0 & 0 & 0 & 0 & 0 \\
 0 & 0 & 0 & 0 & 0 & 0 & 0 & 1 &0 & 0 & 0 & 0 \\
 0 & 0 & 0 & 0 & 0 & 0 & 0 & 0 & 1 & 0 & 0 & 0 \\
 0 & 0 & 0 & 1 & 0 & 0 & 0 & 0 & 0 & 0 & 0 & 0 \\
 0 & 0 & 0 & 0 & 0 & 0 & 0 & 0 & 0 & 0 & 1 & 0 \\
  0 & 0 & 0 & 0 & 0 & 0 & 0 & 0 & 0 & 0 & 0 & 1 \\
\end{array}%
\right),
\end{equation}
the algebra (\ref{T3T3algebra}) is
\begin{eqnarray}
\begin{array}{ll}
[Z_5,Z_6] = mZ_1,& [X^4,Z_6] = -mZ_2,\\
{[X^4,Z_5]} = mZ_3,&[{X}^1,Z_5] = m{X}^6,\\
{[}{X}^2,Z_6] = mZ_4,& [{X}^2,X^4] = -m{X}^6,\\
{[}{X}^3,X^4] = m{X}^5,& [{X}^1,Z_6] = -m{X}^5,\\
{[}{X}^3,Z_5] = -mZ_4.& 
\end{array}
\end{eqnarray}
The $X^m$ generate a subgroup $\tilde{G}$. Taking the quotient by  $\tilde{G}$ gives a T-fold with  metric and $B$-field given by
\begin{eqnarray}
ds^2 &=& (dz^1+mz^6dz^5)^2 +\frac{1}{1+m^2\Big[(z^5)^2+(z^6)^2\Big]}\Big[(dz)^2+(dz^3)^2+(dz^4)^2\Big]\nonumber\\
& & +\frac{1}{1+m^2\Big[(z^5)^2+(z^6)^2\Big]}\Big(mz^5dz^2+mz^6dz^3\Big)^2+(dz^5)^2+(dz^6)^2,\\
B &=& \frac{m}{1+m^2\Big[(z^5)^2+(z^6)^2\Big]}\Big(z^5 dz^3\wedge dz^4-z^6dz^2\wedge dz^4\Big).
\end{eqnarray}

\subsection{$T^3$ bundle over $T^4$}
\label{T3T4}
Consider the seven-dimensional nilpotent Lie algebra whose  only non-vanishing commutators are
\begin{eqnarray}
\begin{array}{ll}
[T_4,T_5] = mT_1,& [T_6,T_7] = mT_1,\\
{[T_4,T_6]} = mT_2,& [T_5,T_7] = -mT_2.\\
{[T_4,T_7]} = mT_3,& [T_5,T_6] = mT_3.
\end{array}
\end{eqnarray}
The corresponding fourteen-dimensional group has a Lie algebra whose only non-zero commutators are
\begin{eqnarray}
\begin{array}{ll}
[T_4,T_5] = mT_1,&[T_6,T_7] = mT_1,\\
{[T_4,T_6]} = mT_2,& [T_5,T_7] = -mT_2.\\
{[T_4,T_7]} = mT_3,&[T_5,T_6] = mT_3,\\
{[}\tilde{T}^1,T_4] = m\tilde{T}^5,& [\tilde{T}^1,T_5] = -m\tilde{T}^4,\\
{[}\tilde{T}^1,T_6] = m\tilde{T}^7,& [\tilde{T}^1,T_7] = -m\tilde{T}^6,\\
{[}\tilde{T}^2,T_4] = m\tilde{T}^6,&[\tilde{T}^2,T_6] = -m\tilde{T}^4,\\
{[}\tilde{T}^2,T_7] = m\tilde{T}^5,& [\tilde{T}^2,T_5] = -m\tilde{T}^7,\\
{[}\tilde{T}^3,T_4] = m\tilde{T}^7,& [\tilde{T}^3,T_7] = -m\tilde{T}^4,\\
{[}\tilde{T}^3,T_5] = m\tilde{T}^6,& [\tilde{T}^3,T_6] = -m\tilde{T}^5.\label{T3T4algebra}
\end{array}
\end{eqnarray}
The left-invariant one-forms are 
\begin{eqnarray}
\begin{array}{ll}
P^1 = dz^1+mz^5dz^4+mz^7dz^6,& Q_1 = d\tilde{z}_1,  \\
P^2 = dz^2+mz^6dz^4-mz^7dz^5, & Q_2 = d\tilde{z}_2,\\
P^3 = dz^3+mz^7dz^4+mz^6dz^5, & Q_3 = d\tilde{z}_3,\\
P^4 = dz^4, & Q_4 = d\tilde{z}_4-mz^5d\tilde{z}_1-mz^6d\tilde{z}_2+mz^4d\tilde{z}_3,\\
P^5 = dz^5, & Q_5 = d\tilde{z}_5+mz^4d\tilde{z}_1+mz^7d\tilde{z}_2-mz^6d\tilde{z}_3,\\
P^6 = dz^6, & Q_6 = d\tilde{z}_6-mz^7d\tilde{z}_1+mz^4d\tilde{z}_2+mz^5d\tilde{z}_3,\\
P^7 = dz^7, & Q_7 = d\tilde{z}_7+mz^6d\tilde{z}_1-mz^5d\tilde{z}_2+mz^4d\tilde{z}_3.
\end{array}
\end{eqnarray}

Choosing the polarisation $\Theta = 1$, the algebra (\ref{T3T4algebra}) is
\begin{eqnarray}
\begin{array}{ll}
[Z_4,Z_5] = mZ_1,&[Z_6,Z_7] = mZ_1,\\
{[Z_4,Z_6]} = mZ_2,& [Z_5,Z_7] = -mZ_2.\\
{[Z_4,Z_7]} = mZ_3,&[Z_5,Z_6] = mZ_3,\\
{[}{X}^1,Z_4] = m{X}^5,& [{X}^1,Z_5] = -m{X}^4,\\
{[}{X}^1,Z_6] = m{X}^7,& [{X}^1,Z_7] = -m{X}^6,\\
{[}{X}^2,Z_4] = m{X}^6,&[{X}^2,Z_6] = -m{X}^4,\\
{[}{X}^2,Z_7] = m{X}^5,& [{X}^2,Z_5] = -m{X}^7,\\
{[}{X}^3,Z_4] = m{X}^7,& [{X}^3,Z_7] = -m{X}^4,\\
{[}{X}^3,Z_5] = m{X}^6,& [{X}^3,T_6] = -m{X}^5.
\end{array}
\end{eqnarray}
The $X^m$ generate an abelian subgroup $\tilde{G}$. Taking the quotient by $\tilde{G}$ gives the metric
\begin{eqnarray}
ds^2 &=& \Big(dz^1+m(z^5dz^4+z^7dz^6)\Big)^2+\Big(dz^2+m(z^6dz^4-z^7dz^5)\Big)^2+\Big(dz^3+m(z^7dz^4+z^6dz^5)\Big)^2\nonumber\\
& & +(dz^4)^2+(dz^5)^2+(dz^6)^2+(dz^7)^2.
\end{eqnarray}

Choosing the polarization
\begin{equation}
\Theta=\left(%
\begin{array}{cccccccccccccc}
  0 & 0 & 0 & 0 & 0 & 0 & 0 & 1 & 0 & 0 & 0 & 0 & 0 & 0 \\
  0 & 0 & 0 & 0 & 0 & 0 & 0 & 0 & 1 & 0 & 0 & 0 & 0 & 0 \\
  0 & 0 & 0 & 0 & 0 & 0 & 0 & 0 & 0 & 1 & 0 & 0 & 0 & 0 \\
  0 & 0 & 0 & 1 & 0 & 0 & 0 & 0 & 0 & 0 & 0 & 0 & 0 & 0 \\
  0 & 0 & 0 & 0 & 1 & 0 & 0 & 0 & 0 & 0 & 0 & 0 & 0 & 0 \\
  0 & 0 & 0 & 0 & 0 & 1 & 0 & 0 & 0 & 0 & 0 & 0 & 0 & 0 \\
  0 & 0 & 0 & 0 & 0 & 0 & 1 & 0 & 0 & 0 & 0 & 0 & 0 & 0 \\
  1 & 0 & 0 & 0 & 0 & 0 & 0 & 0 & 0 & 0 & 0 & 0 & 0 & 0 \\
  0 & 1 & 0 & 0 & 0 & 0 & 0 & 0 & 0 & 0 & 0 & 0 & 0 & 0 \\
  0 & 0 & 1 & 0 & 0 & 0 & 0 & 0 & 0 & 0 & 0 & 0 & 0 & 0 \\
  0 & 0 & 0 & 0 & 0 & 0 & 0 & 0 & 0 & 0 & 1 & 0 & 0 & 0 \\
  0 & 0 & 0 & 0 & 0 & 0 & 0 & 0 & 0 & 0 & 0 & 1 & 0 & 0 \\
  0 & 0 & 0 & 0 & 0 & 0 & 0 & 0 & 0 & 0 & 0 & 0 & 1 & 0 \\
  0 & 0 & 0 & 0 & 0 & 0 & 0 & 0 & 0 & 0 & 0 & 0 & 0 & 1 \\
\end{array}%
\right),
\end{equation}
the algebra (\ref{T3T4algebra}) is
\begin{eqnarray}
\begin{array}{ll}
[Z_4,Z_5] = mX^1,&[Z_6,Z_7] = mX^1,\\
{[Z_4,Z_6]} = mX^2,& [Z_5,Z_7] = -mX^2.\\
{[Z_4,Z_7]} = mX^3,&[Z_5,Z_6] = mX^3,\\
{[}Z_1,Z_4] = m{X}^5,& [Z_1,Z_5] = -m{X}^4,\\
{[}Z_1,Z_6] = m{X}^7,& [Z_1,Z_7] = -m{X}^6,\\
{[}Z_2,Z_4] = m{X}^6,&[Z_2,Z_6] = -m{X}^4,\\
{[}Z_2,Z_7] = m{X}^5,& [Z_2,Z_5] = -m{X}^7,\\
{[}Z_3,Z_4] = m{X}^7,& [Z_3,Z_7] = -m{X}^4,\\
{[}Z_3,Z_5] = m{X}^6,& [Z_3,T_6] = -m{X}^5.
\end{array}
\end{eqnarray}
The $X^m$ generate an abelian subgroup $\tilde{G}$. Taking the quotient by $\tilde{G}$ gives the $T^7$ with $H$-flux.
\begin{equation}
ds^2 = (dz^1)^2+(dz^2)^2+(dz^3)^2+(dz^4)^2+(dz^5)^2+(dz^6)^2+(dz^7)^2,
\end{equation}
\begin{eqnarray}
H =  &=& -mdz^1 \wedge dz^4\wedge dz^5 - mdz^1\wedge dz^6 \wedge dz^7 -mdz^2 \wedge dz^4\wedge dz^6 \nonumber\\
& &- mdz^3\wedge dz^4 \wedge dz^7 - mdz^3 \wedge dz^5 \wedge dz^6+mdz^2 \wedge dz^5\wedge dz^7.
\end{eqnarray}

Choosing the polarization
\begin{equation}
\Theta=\left(%
\begin{array}{cccccccccccccc}
  1 & 0 & 0 & 0 & 0 & 0 & 0 & 0 & 0 & 0 & 0 & 0 & 0 & 0 \\
  0 & 1 & 0 & 0 & 0 & 0 & 0 & 0 & 0 & 0 & 0 & 0 & 0 & 0 \\
  0 & 0 & 1 & 0 & 0 & 0 & 0 & 0 & 0 & 0 & 0 & 0 & 0 & 0 \\
  0 & 0 & 0 & 0 & 0 & 0 & 0 & 0 & 0 & 0 & 1 & 0 & 0 & 0 \\
  0 & 0 & 0 & 0 & 1 & 0 & 0 & 0 & 0 & 0 & 0 & 0 & 0 & 0 \\
  0 & 0 & 0 & 0 & 0 & 1 & 0 & 0 & 0 & 0 & 0 & 0 & 0 & 0 \\
  0 & 0 & 0 & 0 & 0 & 0 & 1 & 0 & 0 & 0 & 0 & 0 & 0 & 0 \\
  0 & 0 & 0 & 0 & 0 & 0 & 0 & 1 & 0 & 0 & 0 & 0 & 0 & 0 \\
  0 & 0 & 0 & 0 & 0 & 0 & 0 & 0 & 1 & 0 & 0 & 0 & 0 & 0 \\
  0 & 0 & 0 & 0 & 0 & 0 & 0 & 0 & 0 & 1 & 0 & 0 & 0 & 0 \\
  0 & 0 & 0 & 1 & 0 & 0 & 0 & 0 & 0 & 0 & 0 & 0 & 0 & 0 \\
  0 & 0 & 0 & 0 & 0 & 0 & 0 & 0 & 0 & 0 & 0 & 1 & 0 & 0 \\
  0 & 0 & 0 & 0 & 0 & 0 & 0 & 0 & 0 & 0 & 0 & 0 & 1 & 0 \\
  0 & 0 & 0 & 0 & 0 & 0 & 0 & 0 & 0 & 0 & 0 & 0 & 0 & 1 \\
\end{array}%
\right),
\end{equation}
the algebra (\ref{T3T4algebra}) is\begin{eqnarray}
\begin{array}{ll}
[X^4,Z_5] = mZ_1,&[Z_6,Z_7] = mZ_1,\\
{[X^4,Z_6]} = mZ_2,& [Z_5,Z_7] = -mZ_2.\\
{[X^4,Z_7]} = mZ_3,&[Z_5,Z_6] = -mZ_3,\\
{[}{X}^1,X^4] = m{X}^5,& [{X}^1,Z_5] = -mZ_4,\\
{[}{X}^1,Z_6] = m{X}^7,& [{X}^1,Z_7] = -m{X}^6,\\
{[}{X}^2,X^4] = m{X}^6,&[{X}^2,Z_6] = -mZ_4,\\
{[}{X}^2,Z_7] = m{X}^5,& [{X}^2,Z_5] = -m{X}^7,\\
{[}{X}^3,X^4] = m{X}^7,& [{X}^3,Z_7] = -mZ_4,\\
{[}{X}^3,Z_5] = m{X}^6,& [{X}^3,T_6] = -m{X}^5,
\end{array}
\end{eqnarray}
The $X^m$ generate a subgroup $\tilde{G}$. Taking the quotient gives the T-fold with the metric and $B$-field given by(\ref{T-fold2metric}), (\ref{T-fold2Bfield}).

\subsection{$S^1$ bundle over $T^6$}
\label{S1T6}
Consider the seven-dimensional nilpotent Lie algebra whose  only non-vanishing commutators are
\begin{eqnarray}
\begin{array}{ll}
[T_2,T_3] = mT_1,& [T_4,T_5] = mT_1,\\
{[T_6,T_7]} = mT_1.&
\end{array}
\end{eqnarray}
The corresponding fourteen-dimensional group has a Lie algebra whose only non-zero commutators are
\begin{eqnarray}
\begin{array}{ll}
[T_2,T_3] = mT_1,& [T_4,T_5] = mT_1,\\
{[T_6,T_7]} = mT_1.& [\tilde{T}^1,T_3] = -m\tilde{T}^2,\\
{[}\tilde{T}^1,T_2] = m\tilde{T}^3,& [\tilde{T}^1,T_5] = -m\tilde{T}^4,\\
{[}\tilde{T}^1,T_4] = m\tilde{T}^5,& [\tilde{T}^1,T_7] = -m\tilde{T}^6,\\
{[}\tilde{T}^1,T_6] = m\tilde{T}^7.&\label{T1T6algebra}
\end{array}
\end{eqnarray}
The left-invariant one-forms are 
\begin{eqnarray}
\begin{array}{ll}
P^1 = dz^1+mz^3dz^2+mz^5dz^4+mz^7dz^6,& Q_1 = d\tilde{z}_1,  \\
P^2 = dz^2, & Q_2 = d\tilde{z}_2-mz^3d\tilde{z}_1,\\
P^3 = dz^3, & Q_3 = d\tilde{z}_3+mz^2d\tilde{z}_1,\\
P^4 = dz^4, & Q_4 = d\tilde{z}_4-mz^5d\tilde{z}_1,\\
P^5 = dz^5, & Q_5 = d\tilde{z}_5+mz^4d\tilde{z}_1,\\
P^6 = dz^6, & Q_6 = d\tilde{z}_6-mz^7d\tilde{z}_1,\\
P^7 = dz^7, & Q_7 = d\tilde{z}_7+mz^6d\tilde{z}_1.
\end{array}
\end{eqnarray}

Choosing the polarisation $\Theta = 1$, the algebra (\ref{T1T6algebra}) is
\begin{eqnarray}
\begin{array}{ll}
[Z_2,Z_3] = mZ_1,& [Z_4,Z_5] = mZ_1,\\
{[Z_6,Z_7]} = mZ_1.& [{X}^1,Z_3] = -m{X}^2,\\
{[}{X}^1,Z_2] = m{X}^3,& [{X}^1,Z_5] = -m{X}^4,\\
{[}{X}^1,Z_4] = m{X}^5,& [{X}^1,Z_7] = -m{X}^6,\\
{[}{X}^1,Z_6] = m{X}^7.&
\end{array}
\end{eqnarray}
The $X^m$ generate an abelian subgroup $\tilde{G}$. Taking the quotient by $\tilde{G}$ gives the metric
\begin{equation}
ds^2 = \Big(dz^1+m(z^3dz^2+z^5dz^4+z^7dz^6)\Big)^2+(dz^2)^2+(dz^3)^2+(dz^4)^2+(dz^5)^2+(dz^6)^2+(dz^7)^2.
\end{equation}

Choosing the polarization
\begin{equation}
\Theta=\left(%
\begin{array}{cccccccccccccc}
  0 & 0 & 0 & 0 & 0 & 0 & 0 & 1 & 0 & 0 & 0 & 0 & 0 & 0 \\
  0 & 1 & 0 & 0 & 0 & 0 & 0 & 0 & 0 & 0 & 0 & 0 & 0 & 0 \\
  0 & 0 & 1 & 0 & 0 & 0 & 0 & 0 & 0 & 0 & 0 & 0 & 0 & 0 \\
  0 & 0 & 0 & 1 & 0 & 0 & 0 & 0 & 0 & 0 & 0 & 0 & 0 & 0 \\
  0 & 0 & 0 & 0 & 1 & 0 & 0 & 0 & 0 & 0 & 0 & 0 & 0 & 0 \\
  0 & 0 & 0 & 0 & 0 & 1 & 0 & 0 & 0 & 0 & 0 & 0 & 0 & 0 \\
  0 & 0 & 0 & 0 & 0 & 0 & 1 & 0 & 0 & 0 & 0 & 0 & 0 & 0 \\
  1 & 0 & 0 & 0 & 0 & 0 & 0 & 0 & 0 & 0 & 0 & 0 & 0 & 0 \\
  0 & 0 & 0 & 0 & 0 & 0 & 0 & 0 & 1 & 0 & 0 & 0 & 0 & 0 \\
  0 & 0 & 0 & 0 & 0 & 0 & 0 & 0 & 0 & 1 & 0 & 0 & 0 & 0 \\
  0 & 0 & 0 & 0 & 0 & 0 & 0 & 0 & 0 & 0 & 1 & 0 & 0 & 0 \\
  0 & 0 & 0 & 0 & 0 & 0 & 0 & 0 & 0 & 0 & 0 & 1 & 0 & 0 \\
  0 & 0 & 0 & 0 & 0 & 0 & 0 & 0 & 0 & 0 & 0 & 0 & 1 & 0 \\
  0 & 0 & 0 & 0 & 0 & 0 & 0 & 0 & 0 & 0 & 0 & 0 & 0 & 1 \\
\end{array}%
\right),
\end{equation}
the algebra (\ref{T1T6algebra}) is
\begin{eqnarray}
\begin{array}{ll}
[Z_2,Z_3] = mX^1,& [Z_4,Z_5] = mX^1,\\
{[Z_6,Z_7]} = mX^1.& [Z_1,Z_3] = -m{X}^2,\\
{[}Z_1,Z_2] = m{X}^3,& [Z_1,Z_5] = -m{X}^4,\\
{[}Z_1,Z_4] = m{X}^5,& [Z_1,Z_7] = -m{X}^6,\\
{[}Z_1,Z_6] = m{X}^7.&
\end{array}
\end{eqnarray}
The $X^m$ generate an abelian subgroup $\tilde{G}$. Taking the quotient gives the $T^7$ with $H$-flux.
\begin{equation}
ds^2 = (dz^1)^2+(dz^2)^2+(dz^3)^2+(dz^4)^2+(dz^5)^2+(dz^6)^2+(dz^7)^2,
\end{equation}
\begin{equation}
H =  -mdz^1 \wedge dz^2\wedge dz^3 - mdz^1\wedge dz^4 \wedge dz^5 -mdz^1 \wedge dz^6\wedge dz^7.
\end{equation}

Choosing the polarization
\begin{equation}
\Theta=\left(%
\begin{array}{cccccccccccccc}
  1 & 0 & 0 & 0 & 0 & 0 & 0 & 0 & 0 & 0 & 0 & 0 & 0 & 0 \\
  0 & 0 & 0 & 0 & 0 & 0 & 0 & 0 & 1 & 0 & 0 & 0 & 0 & 0 \\
  0 & 0 & 1 & 0 & 0 & 0 & 0 & 0 & 0 & 0 & 0 & 0 & 0 & 0 \\
  0 & 0 & 0 & 0 & 0 & 0 & 0 & 0 & 0 & 0 & 1 & 0 & 0 & 0 \\
  0 & 0 & 0 & 0 & 1 & 0 & 0 & 0 & 0 & 0 & 0 & 0 & 0 & 0 \\
  0 & 0 & 0 & 0 & 0 & 0 & 0 & 0 & 0 & 0 & 0 & 0 & 1 & 0 \\
  0 & 0 & 0 & 0 & 0 & 0 & 1 & 0 & 0 & 0 & 0 & 0 & 0 & 0 \\
  0 & 0 & 0 & 0 & 0 & 0 & 0 & 1 & 0 & 0 & 0 & 0 & 0 & 0 \\
  0 & 1 & 0 & 0 & 0 & 0 & 0 & 0 & 0 & 0 & 0 & 0 & 0 & 0 \\
  0 & 0 & 0 & 0 & 0 & 0 & 0 & 0 & 0 & 1 & 0 & 0 & 0 & 0 \\
  0 & 0 & 0 & 1 & 0 & 0 & 0 & 0 & 0 & 0 & 0 & 0 & 0 & 0 \\
  0 & 0 & 0 & 0 & 0 & 0 & 0 & 0 & 0 & 0 & 0 & 1 & 0 & 0 \\
  0 & 0 & 0 & 0 & 0 & 1 & 0 & 0 & 0 & 0 & 0 & 0 & 0 & 0 \\
  0 & 0 & 0 & 0 & 0 & 0 & 0 & 0 & 0 & 0 & 0 & 0 & 0 & 1 \\
\end{array}%
\right),
\end{equation}
the algebra (\ref{T1T6algebra}) is
\begin{eqnarray}
\begin{array}{ll}
[X^2,Z_3] = mZ_1,& [X^4,Z_5] = mZ_1,\\
{[X^6,Z_7]} = mZ_1.& [{X}^1,Z_3] = -mZ_2,\\
{[}{X}^1,X^2] = m{X}^3,& [{X}^1,Z_5] = -mZ_4,\\
{[}{X}^1,X^4] = m{X}^5,& [{X}^1,Z_7] = -mZ_6,\\
{[}{X}^1,X^6] = m{X}^7.&
\end{array}
\end{eqnarray}
The $X^m$ generate a subgroup $\tilde{G}$. Taking the quotient gives the T-fold with the metric and $B$-field given by
\begin{eqnarray}
ds^2 &=& \frac{1}{1+m^2\Big[(z^3)^2+(z^5)^2+(z^7)^2\Big]}\Big[(dz^1)^2+(dz^2)^2+(dz^4)^2+(dz^6)^2\Big]\nonumber\\
& &+(dz^3)^2+(dz^5)^2+(dz^7)^2+\frac{1}{1+m^2\Big[(z^3)^2+(z^5)^2+(z^7)^2\Big]}\Big(mz^5dz^2-mz^3dz^4\Big)^2\nonumber\\
& &+\frac{1}{1+m^2\Big[(z^3)^2+(z^5)^2+(z^7)^2\Big]}\Big(mz^7dz^2-mz^3dz^6\Big)^2\nonumber\\
& &+\frac{1}{1+m^2\Big[(z^3)^2+(z^5)^2+(z^7)^2\Big]}\Big(mz^7dz^4-mz^5dz^6\Big)^2,
\end{eqnarray}
\begin{eqnarray}
B = \frac{m}{1+m^2\Big[(z^3)^2+(z^5)^2+(z^7)^2\Big]}\Big(z^3dz^1\wedge dz^2+z^5dz^1\wedge dz^4+z^7dz^1\wedge dz^6\Big).
\end{eqnarray}

\section{Nilmanifolds Fibred over a Line.} 

\subsection{The Doubled Sigma model for Nilmanifolds Fibred over a Line}

In the previous section, the doubled geometries of various nilmanifolds  were constructed. In this section, we will construct the doubled geometries of the 
corresponding  special holonomy spaces. 
The special holonomy  solutions are of
 the form $\mathcal{N}\times \mathbb{R}$, where $\mathcal{N}$ is the nilmanifold. The doubled formulation of this space is constructed by doubling $d$-dimensional nilmanifold $\mathcal{N}$ to the $2d$-dimensional nilmanifold $\mathcal{M}$. That is,  the space $\mathcal{N}\times \mathbb{R}$ is extended to $\mathcal{M}\times \mathbb{R}$.

Consider the non-linear sigma model for the   special holonomy space $\mathcal{N}\times\mathbb{R}$
\begin{equation}
S_{\mathcal{N}\times\mathbb{R}} = \frac{1}{2}\oint_{\Sigma}\Big(V^p(\tau)d\tau\wedge *d\tau+x_{mn}(\tau)P^m\wedge *P^n\Big),
\end{equation}
where $V(\tau)$ is a piecewise linear function on the line with coordinate $\tau$, $p$ is an integer  depending on the choice of nilmanifold, $x_{mn}$ is a symmetric matrix whose non-zero entries are powers of $V(\tau)$, and $P^m$ are the left-invariant one-forms on $\mathcal{N}$. This action can be generalized to the non-linear sigma model of $\mathcal{M}\times \mathbb{R}$, which is given by
\begin{equation}
S_{\mathcal{M}\times\mathbb{R}} = \frac{1}{2}\oint_{\Sigma}V^p(\tau)d\tau\wedge *d\tau+\frac{1}{4}\oint_\Sigma \mathcal{M}_{MN}(\tau)\hat{\mathcal{P}}^M\wedge*\hat{\mathcal{P}}^N+\frac{1}{2}\int_S \mathcal{K},\label{DoubleAction}
\end{equation}
where $\mathcal{M}_{MN}(\tau)$ is a symmetric matrix whose non-zero entries are powers of $V(\tau)$, $\hat{\mathcal{P}}^M$ is  the pull-back to $\Sigma$ of the left-invariant one-forms on $\mathcal{M}$ to $\Sigma$, and $\mathcal{K}$ is the 3-form on $\mathcal{M}$   pulled back
to $S$ defined in (\ref{Kthreeform}).
This non-linear sigma model contains all the T-duals  of  the nilmanifold $\mathcal{N}$. Different T-dual backgrounds are selected by choosing different  polarizations as in  section \ref{HigherNilfold}.

In the next subsection, we discuss the  example of the 3-dimensional nilfold in some detail.
Then in the following subsection, we give the models for the other nilmanifolds; the details of these are similar to those of the 3-dimensional nilfold.

\subsection{Example: 3-dimensional Nilmanifold}
The doubled space for the 3-dimensional nilmanifold $\mathcal{N}$ is given by a 6-dimensional nilmanifold $\mathcal{M}$. The non-vanishing commutation relations are 
\begin{equation}
[T_x,T_z]=mT_y,  \qquad   
\qquad  [T_x,\tilde T ^y]=mX^z,  \qquad  [T_z,\tilde T^y]=-mX^x.
\end{equation}
The left-invariant one-forms on this space are
\begin{eqnarray}
\begin{array}{lll}
P^x = dx,& P^y = dy-mxdz, & P^z = dz,  \\
Q_x = d\tilde{x}-mzd\tilde{y},&Q_y = d\tilde{y} ,& Q_z = d\tilde{z}+mxd\tilde{y}.
\end{array}
\end{eqnarray}
In this case, the doubled non-linear sigma model on $\mathcal{M}\times \mathbb{R}$ is
\begin{equation}
S_{\mathcal{M}\times\mathbb{R}} = \frac{1}{2}\oint_{\Sigma}V(\tau)d\tau\wedge *d\tau+\frac{1}{4}\oint_\Sigma \mathcal{M}_{MN}(\tau)\hat{\mathcal{P}}^M\wedge*\hat{\mathcal{P}}^N+\frac{1}{2}\int_S \mathcal{K},\label{SMtimesR}
\end{equation}
 where $\mathcal{M}_{MN}(\tau)$ is given by 
 \begin{equation}
\mathcal{M}_{MN}(\tau) =
 \left(
\begin{array}{cccccc}
V(\tau) & 0 & 0 & 0 & 0 & 0 \\
0 & 1/V(\tau) & 0 & 0 & 0 & 0 \\
0 & 0 & V(\tau) & 0 & 0 & 0 \\
0 & 0 & 0 & 1/V(\tau) & 0 & 0 \\
0 & 0 & 0 & 0 & V(\tau) & 0 \\
0 & 0 & 0 & 0 & 0 & 1/V(\tau)
\end{array}\right).
\end{equation}
To obtain a nilfold bundle over a line, the polarization is chosen as $\Theta = 1$.
In this case, the generalised metric (\ref{generalizemetric}), which is the second term in (\ref{SMtimesR}), is given by
\begin{equation}
\mathcal{H} = \frac{1}{2}\mathcal{M}_{MN}(\tau)\mathcal{P}^M\otimes\mathcal{P}^N = \frac{1}{2}\mathcal{H}_{MN}(\tau,x^m)\Phi^M\otimes\Phi^N.
\end{equation} 
With this polarization tensor, we define $\mathcal{H}_{\hat{M}\hat{N}}(\tau,x^m)$ as in (\ref{hatGenmetric}). Its components can be used to define $g_{mn}$ and $B_{mn}$ as in (\ref{H}).
The metric (\ref{metricmn}) and the $H$-flux (\ref{H conjecture})   obtained in this way are
\begin{eqnarray}
ds^2_3 &=& V(\tau)\Big(dx^2+dz^2\Big)+\frac{1}{V(\tau)}(dy-mxdz)^2,\\
H &=& 0.
\end{eqnarray}
This results in the metric
\begin{equation}
ds^2 = V(\tau)\Big((d\tau)^2+(dx)^2+(dz)^2\Big)+\frac{1}{V(\tau)}(dy-mxdz)^2.
\end{equation} 

To obtain a $T^3$ with $H$-flux, the polarization  tensor is chosen as 
$$
\Theta=\left(%
\begin{array}{cccccc}
  1 & 0 & 0 & 0 & 0 & 0 \\
  0 & 0 & 0 & 0 & 1 & 0 \\
  0 & 0 & 1 & 0 & 0 & 0 \\
  0 & 0 & 0 & 1 & 0 & 0 \\
  0 & 1 & 0 & 0 & 0 & 0 \\
  0 & 0 & 0 & 0 & 0 & 1 \\
\end{array}
\right).
$$
In this polarization, the matrix of scalar moduli is given by
\begin{equation}
\mathcal{M}_{MN}(\tau) =
 \left(
\begin{array}{cccccc}
V(\tau) & 0 & 0 & 0 & 0 & 0 \\
0 & V(\tau) & 0 & 0 & 0 & 0 \\
0 & 0 & V(\tau) & 0 & 0 & 0 \\
0 & 0 & 0 & 1/V(\tau) & 0 & 0 \\
0 & 0 & 0 & 0 & 1/V(\tau) & 0 \\
0 & 0 & 0 & 0 & 0 & 1/V(\tau)
\end{array}\right)
\end{equation}
and the left-invariant one-forms are 
\begin{eqnarray}
\begin{array}{lll}
P^x = dx,& P^y = dy, & P^z = dz,  \\
Q_x = d\tilde{x}-mzd{y},&Q_y = d\tilde{y}-mxdz, & Q_z = d\tilde{z}+mxd{y}.
\end{array}
\end{eqnarray}
With this polarization tensor, we define $\mathcal{H}_{\hat{M}\hat{N}}(\tau,x^m)$ by(\ref{hatGenmetric}). Its components can be used to define $g_{mn}$ and $B_{mn}$ as in (\ref{H}).
The  resulting metric (\ref{metricmn}) and the $H$-flux (\ref{H conjecture}) are
\begin{eqnarray}
ds^2_3 &=& V(\tau)\Big(dx^2+dy^2+dz^2\Big)\\
H &=& mdx\wedge dy \wedge dz.
\end{eqnarray}
This results in the metric and three-form 
\begin{eqnarray}
ds^2 &=& V(\tau)\Big((d\tau)^2+(dx)^2+(dy)^2+(dz)^2\Big),\\
H &=& mdx\wedge dy\wedge dz.
\end{eqnarray}

To obtain the  T-fold background, the polarization is chosen as 
$$
\Theta=\left(%
\begin{array}{cccccc}
  1 & 0 & 0 & 0 & 0 & 0 \\
  0 & 1 & 0 & 0 & 0 & 0 \\
  0 & 0 & 0 & 0 & 0 & 1 \\
  0 & 0 & 0 & 1 & 0 & 0 \\
  0 & 0 & 0 & 0 & 1 & 0 \\
  0 & 0 & 1 & 0 & 0 & 0 \\
\end{array}%
\right).
$$
In this polarization, \begin{equation}
\mathcal{M}_{MN}(\tau) =
 \left(
\begin{array}{cccccc}
V(\tau) & 0 & 0 & 0 & 0 & 0 \\
0 & 1/V(\tau) & 0 & 0 & 0 & 0 \\
0 & 0 & 1/V(\tau) & 0 & 0 & 0 \\
0 & 0 & 0 & 1/V(\tau) & 0 & 0 \\
0 & 0 & 0 & 0 & V(\tau) & 0 \\
0 & 0 & 0 & 0 & 0 & V(\tau)
\end{array}\right).
\end{equation}
and the left-invariant one-forms are
\begin{eqnarray}
\begin{array}{lll}
P^x = dx,& P^y = dy-mxd\tilde{z}, & P^z = dz+mxd\tilde{y},  \\
Q_x = d\tilde{x}-m\tilde{z}d\tilde{y},&Q_y = d\tilde{y}, & Q_z = d\tilde{z}.
\end{array}
\end{eqnarray}
We then define $\mathcal{H}_{\hat{M}\hat{N}}(\tau,x^m)$ as in (\ref{hatGenmetric}). Its components can be used to define $g_{mn}$ and $B_{mn}$ as in (\ref{H}).
This gives the metric (\ref{metricmn}) and the $B$-field from the $H$-flux (\ref{H conjecture}) can be obtained
\begin{eqnarray}
ds^2 &=& V(\tau)\Big((d\tau)^2+(dx)^2\Big)+\frac{V(\tau)}{V^2(\tau)+(mx)^2}\Big((dy)^2+(dz)^2\Big),\\
B &=& \frac{-mx}{V^2(\tau)+(mx)^2}dy\wedge dz.
\end{eqnarray}

 Choosing the polarization (\ref{RFLUXsection}) gives the {R}-flux background. In this case, no local 4-dimensional field configuration can be extracted from the 7-dimensional doubled space.
This construction can be generalized to the higher dimensional nilmanifolds.

\subsection{The Special Holonomy Nilmanifolds Fibred over a Line}
\subsubsection{$S^1$ bundle over $T^4$ and $T^2$ bundle over $T^3$}
The doubled Lie algebras  (\ref{T1T4algebra}) and (\ref{T2T3algebra}) are isomorphic. This means 
that they define the same doubled space, and reflects the fact that the two corresponding undoubled nilmanifolds are T-dual.

The doubled geometry of $ \mathcal{M}$
is given in subsection \ref{S1T4}.
The 
 doubled non-linear sigma model on $\mathcal{M}\times \mathbb{R}$ is
\begin{eqnarray}
S_{\mathcal{M}\times\mathbb{R}} = \frac{1}{2}\oint_{\Sigma}V^2(\tau)d\tau\wedge *d\tau+\frac{1}{4}\oint_\Sigma \mathcal{M}_{MN}(\tau)\hat{\mathcal{P}}^M\wedge*\hat{\mathcal{P}}^N+\frac{1}{2}\int_V \mathcal{K}
\end{eqnarray}
 where $\mathcal{M}_{MN}(\tau)$, in the basis given in section \ref{S1T4}, is given by 
\begin{equation}
\mathcal{M}_{MN}(\tau) =
 \left(
\begin{array}{cccccccccc}
V^{-2} & 0 & 0 & 0 & 0 & 0 & 0 & 0 & 0 & 0\\
0 & V & 0 & 0 & 0 & 0 & 0 & 0 & 0 & 0 \\
0 & 0 & V & 0 & 0 & 0& 0 & 0 & 0 & 0 \\
0 & 0 & 0 & V & 0 & 0 & 0 & 0 & 0 & 0\\
0 & 0 & 0 & 0 & V & 0& 0 & 0 & 0 & 0 \\
0 & 0 & 0 & 0 & 0 & V^2& 0 & 0 & 0 & 0 \\
0 & 0 & 0 & 0 & 0 & 0& V^{-1} & 0 & 0 & 0 \\
0 & 0 & 0 & 0 & 0 & 0& 0 & V^{-1} & 0 & 0 \\
0 & 0 & 0 & 0 & 0 & 0& 0 & 0 & V^{-1} & 0 \\
0 & 0 & 0 & 0 & 0 & 0& 0 & 0 & 0 & V^{-1}
\end{array}\right).
\end{equation}

To obtain a 5-dimensional nilmanifold bundle over a line, the polarization is chosen as $\Theta = 1$. With this polarization tensor, we define $\mathcal{H}_{\hat{M}\hat{N}}(\tau,x^m)$ as in (\ref{hatGenmetric}). Its components 
give
$g_{mn}$ and $B_{mn}$ through (\ref{H}), resulting in
\begin{eqnarray}
g &=& \frac{1}{V^2}\big(dz^1+mz^3dz^2+mz^5dz^4\big)^2+V\big((dz^2)^2+(dz^3)^2+(dz^4)^2+(dz^5)^2\big)\\
B &=& 0.
\end{eqnarray}
Then
\begin{eqnarray}
H = dB-\frac{1}{2}d(r^m\wedge q_m) +\frac{1}{2}\mathcal{K},
\end{eqnarray}
where
\begin{eqnarray}
d(r^m\wedge q_m) &=& md\tilde{z}_1\wedge dz^2 \wedge dz^3 +md\tilde{z}_1 \wedge dz^4 \wedge dz^5,\\
 \mathcal{K} &=& m d\tilde{z}_1\wedge dz^2 \wedge dz^3 + md\tilde{z}_1 \wedge dz^4\wedge dz^5
\end{eqnarray}
gives $H=0$.

To obtain a $T^5$ with $H$-flux, the polarization is chosen as
\begin{equation}
\Theta=\left(%
\begin{array}{cccccccccc}
  0 & 0 & 0 & 0 & 0 & 1 & 0 & 0 & 0 & 0  \\
  0 & 1 & 0 & 0 & 0 & 0 & 0 & 0 & 0 & 0  \\
  0 & 0 & 1 & 0 & 0 & 0 & 0 & 0 & 0 & 0  \\
  0 & 0 & 0 & 1 & 0 & 0 & 0 & 0 & 0 & 0  \\
  0 & 0 & 0 & 0 & 1 & 0 & 0 & 0 & 0 & 0  \\
  1 & 0 & 0 & 0 & 0 & 0 & 0 & 0 & 0 & 0  \\
  0 & 0 & 0 & 0 & 0 & 0 & 1 & 0 & 0 & 0  \\
  0 & 0 & 0 & 0 & 0 & 0 & 0 & 1 & 0 & 0  \\
  0 & 0 & 0 & 0 & 0 & 0 & 0 & 0 & 1 & 0  \\
  0 & 0 & 0 & 0 & 0 & 0 & 0 & 0 & 0 & 1  \\
\end{array}%
\right).
\end{equation}
The metric and $H$-flux are then given by
\begin{eqnarray}
g = {V^2}(dz^1)^2+V\big((dz^2)^2+(dz^3)^2+(dz^4)^2+(dz^5)^2\big)
\end{eqnarray}
\begin{eqnarray}
H = -mdz^1\wedge dz^2 \wedge dz^3 - mdz^1\wedge dz^2\wedge dz^5.
\end{eqnarray}

To obtain a $T^2$ bundle over $T^3$, the polarization is chosen as
\begin{equation}
\Theta=\left(%
\begin{array}{cccccccccc}
  0 & 0 & 0 & 0 & 0 & 1 & 0 & 0 & 0 & 0  \\
  0 & 0 & 0 & 0 & 0 & 0 & 1 & 0 & 0 & 0  \\
  0 & 0 & 1 & 0 & 0 & 0 & 0 & 0 & 0 & 0  \\
  0 & 0 & 0 & 0 & 0 & 0 & 0 & 0 & 1 & 0  \\
  0 & 0 & 0 & 0 & 1 & 0 & 0 & 0 & 0 & 0  \\
  1 & 0 & 0 & 0 & 0 & 0 & 0 & 0 & 0 & 0  \\
  0 & 1 & 0 & 0 & 0 & 0 & 0 & 0 & 0 & 0  \\
  0 & 0 & 0 & 0 & 0 & 0 & 0 & 1 & 0 & 0  \\
  0 & 0 & 0 & 1 & 0 & 0 & 0 & 0 & 0 & 0  \\
  0 & 0 & 0 & 0 & 0 & 0 & 0 & 0 & 0 & 1  \\
\end{array}%
\right).
\end{equation}
In this polarization, the $\mathcal{M}_{MN}$ are given by
\begin{equation}
\mathcal{M}_{MN}(\tau) =
 \left(
\begin{array}{cccccccccc}
V^{2} & 0 & 0 & 0 & 0 & 0 & 0 & 0 & 0 & 0\\
0 & V^{-1} & 0 & 0 & 0 & 0 & 0 & 0 & 0 & 0 \\
0 & 0 & V^{-1} & 0 & 0 & 0& 0 & 0 & 0 & 0 \\
0 & 0 & 0 & V & 0 & 0 & 0 & 0 & 0 & 0\\
0 & 0 & 0 & 0 & V & 0& 0 & 0 & 0 & 0 \\
0 & 0 & 0 & 0 & 0 & V^{-2}& 0 & 0 & 0 & 0 \\
0 & 0 & 0 & 0 & 0 & 0& V & 0 & 0 & 0 \\
0 & 0 & 0 & 0 & 0 & 0& 0 & V& 0 & 0 \\
0 & 0 & 0 & 0 & 0 & 0& 0 & 0 & V^{-1} & 0 \\
0 & 0 & 0 & 0 & 0 & 0& 0 & 0 & 0 & V^{-1}
\end{array}\right).
\end{equation}
In this polarization, the metric and $H$-flux are given by
\begin{eqnarray}
g = {V^2}(dz^1)^2+V\big((dz^3)^2+(dz^5)^2\big)+\frac{1}{V}\big((dz^2-mz^3dz^1)^2+(dz^4-mz^5dz^1)^2\big)
\end{eqnarray}


\subsubsection{$T^2$ bundle over $T^4$}
The doubled geometry of $ \mathcal{M}$
is given in subsection \ref{T2T4}.
The 
 doubled non-linear sigma model on $\mathcal{M}\times \mathbb{R}$ is then
 \begin{eqnarray}
S_{\mathcal{M}\times\mathbb{R}} = \frac{1}{2}\oint_{\Sigma}V^4(\tau)d\tau\wedge *d\tau+\frac{1}{4}\oint_\Sigma \mathcal{M}_{MN}(\tau)\hat{\mathcal{P}}^M\wedge*\hat{\mathcal{P}}^N+\frac{1}{2}\int_V \mathcal{K}
\end{eqnarray}
 where $\mathcal{M}_{MN}(\tau)$ is given by 
\begin{equation}
\mathcal{M}_{MN}(\tau) =
 \left(
\begin{array}{cccccccccccc}
V^{-2} & 0 & 0 & 0 & 0 & 0 & 0 & 0 & 0 & 0 & 0 & 0\\
0 & V^{-2} & 0 & 0 & 0 & 0 & 0 & 0 & 0 & 0 & 0 & 0\\
0 & 0 & V^{2} & 0 & 0 & 0& 0 & 0 & 0 & 0 & 0 & 0\\
0 & 0 & 0 & V^{2} & 0 & 0 & 0 & 0 & 0 & 0& 0 & 0\\
0 & 0 & 0 & 0 & V^{2} & 0& 0 & 0 & 0 & 0 & 0 & 0\\
0 & 0 & 0 & 0 & 0 & V^{2}& 0 & 0 & 0 & 0 & 0 & 0\\
0 & 0 & 0 & 0 & 0 & 0& V^{2} & 0 & 0 & 0 & 0 & 0\\
0 & 0 & 0 & 0 & 0 & 0& 0 & V^{2} & 0 & 0& 0 & 0 \\
0 & 0 & 0 & 0 & 0 & 0& 0 & 0 & V^{-2} & 0 & 0 & 0\\
0 & 0 & 0 & 0 & 0 & 0& 0 & 0 & 0 & V^{-2} & 0 & 0\\
0 & 0 & 0 & 0 & 0 & 0& 0 & 0 & 0 & 0 & V^{-2} & 0\\
0 & 0 & 0 & 0 & 0 & 0& 0 & 0 & 0 & 0 & 0 & V^{-2}
\end{array}\right).
\end{equation}

\subsubsection{$T^3$ bundle over $T^3$}
The doubled geometry of $ \mathcal{M}$
is given in subsection \ref{T3T3}.
The 
 doubled non-linear sigma model on $\mathcal{M}\times \mathbb{R}$ is
 \begin{eqnarray}
S_{\mathcal{M}\times\mathbb{R}} = \frac{1}{2}\oint_{\Sigma}V^3(\tau)d\tau\wedge *d\tau+\frac{1}{4}\oint_\Sigma \mathcal{M}_{MN}(\tau)\hat{\mathcal{P}}^M\wedge*\hat{\mathcal{P}}^N+\frac{1}{2}\int_V \mathcal{K}
\end{eqnarray}
 where $\mathcal{M}_{MN}(\tau)$ is given by 
\begin{equation}
\mathcal{M}_{MN}(\tau) =
 \left(
\begin{array}{cccccccccccc}
V^{-1} & 0 & 0 & 0 & 0 & 0 & 0 & 0 & 0 & 0 & 0 & 0\\
0 & V^{-1} & 0 & 0 & 0 & 0 & 0 & 0 & 0 & 0 & 0 & 0\\
0 & 0 & V^{-1} & 0 & 0 & 0& 0 & 0 & 0 & 0 & 0 & 0\\
0 & 0 & 0 & V^{2} & 0 & 0 & 0 & 0 & 0 & 0& 0 & 0\\
0 & 0 & 0 & 0 & V^{2} & 0& 0 & 0 & 0 & 0 & 0 & 0\\
0 & 0 & 0 & 0 & 0 & V^{2}& 0 & 0 & 0 & 0 & 0 & 0\\
0 & 0 & 0 & 0 & 0 & 0& V & 0 & 0 & 0 & 0 & 0\\
0 & 0 & 0 & 0 & 0 & 0& 0 & V & 0 & 0& 0 & 0 \\
0 & 0 & 0 & 0 & 0 & 0& 0 & 0 & V & 0 & 0 & 0\\
0 & 0 & 0 & 0 & 0 & 0& 0 & 0 & 0 & V^{-2} & 0 & 0\\
0 & 0 & 0 & 0 & 0 & 0& 0 & 0 & 0 & 0 & V^{-2} & 0\\
0 & 0 & 0 & 0 & 0 & 0& 0 & 0 & 0 & 0 & 0 & V^{-2}
\end{array}\right).
\end{equation}

\subsubsection{$T^3$ bundle over $T^4$}
The doubled geometry of $ \mathcal{M}$
is given in subsection \ref{T3T4}.
The 
 doubled non-linear sigma model on $\mathcal{M}\times \mathbb{R}$ is
 \begin{eqnarray}
S_{\mathcal{M}\times\mathbb{R}} = \frac{1}{2}\oint_{\Sigma}V^6(\tau)d\tau\wedge *d\tau+\frac{1}{4}\oint_\Sigma \mathcal{M}_{MN}(\tau)\hat{\mathcal{P}}^M\wedge*\hat{\mathcal{P}}^N+\frac{1}{2}\int_V \mathcal{K}
\end{eqnarray} 
where $\mathcal{M}_{MN}(\tau)$ is given by 
\begin{equation}
\mathcal{M}_{MN}(\tau) =
 \left(
\begin{array}{cccccccccccccc}
V^{-2} & 0 & 0 & 0 & 0 & 0 & 0 & 0 & 0 & 0 & 0 & 0 & 0 & 0\\
0 & V^{-2} & 0 & 0 & 0 & 0 & 0 & 0 & 0 & 0 & 0 & 0& 0 & 0\\
0 & 0 & V^{-2} & 0 & 0 & 0& 0 & 0 & 0 & 0 & 0 & 0& 0 & 0\\
0 & 0 & 0 & V^{3} & 0 & 0 & 0 & 0 & 0 & 0& 0 & 0& 0 & 0\\
0 & 0 & 0 & 0 & V^{3} & 0& 0 & 0 & 0 & 0 & 0 & 0& 0 & 0\\
0 & 0 & 0 & 0 & 0 & V^{3}& 0 & 0 & 0 & 0 & 0 & 0& 0 & 0\\
0 & 0 & 0 & 0 & 0 & 0& V^{3} & 0 & 0 & 0 & 0 & 0& 0 & 0\\
0 & 0 & 0 & 0 & 0 & 0& 0 & V^{2} & 0 & 0& 0 & 0 & 0 & 0\\
0 & 0 & 0 & 0 & 0 & 0& 0 & 0 & V^2 & 0 & 0 & 0& 0 & 0\\
0 & 0 & 0 & 0 & 0 & 0& 0 & 0 & 0 & V^{2} & 0 & 0& 0 & 0\\
0 & 0 & 0 & 0 & 0 & 0& 0 & 0 & 0 & 0 & V^{-3} & 0& 0 & 0\\
0 & 0 & 0 & 0 & 0 & 0& 0 & 0 & 0 & 0 & 0 & V^{-3}& 0 & 0 \\
0 & 0 & 0 & 0 & 0 & 0& 0 & 0 & 0 & 0 & 0 & 0& V^{-3} & 0\\
0 & 0 & 0 & 0 & 0 & 0& 0 & 0 & 0 & 0 & 0 & 0& 0 & V^{-3} 
\end{array}\right).
\end{equation}

\subsubsection{$S^1$ bundle over $T^6$}
The doubled geometry of $ \mathcal{M}$
is given in subsection \ref{S1T6}.
The 
 doubled non-linear sigma model on $\mathcal{M}\times \mathbb{R}$ is
 \begin{eqnarray}
S_{\mathcal{M}\times\mathbb{R}} = \frac{1}{2}\oint_{\Sigma}V^3(\tau)d\tau\wedge *d\tau+\frac{1}{4}\oint_\Sigma \mathcal{M}_{MN}(\tau)\hat{\mathcal{P}}^M\wedge*\hat{\mathcal{P}}^N+\frac{1}{2}\int_V \mathcal{K}
\end{eqnarray}
 where $\mathcal{M}_{MN}(\tau)$ is given by 
\begin{equation}
\mathcal{M}_{MN}(\tau) =
 \left(
\begin{array}{cccccccccccccc}
V^{-3} & 0 & 0 & 0 & 0 & 0 & 0 & 0 & 0 & 0 & 0 & 0 & 0 & 0\\
0 & V & 0 & 0 & 0 & 0 & 0 & 0 & 0 & 0 & 0 & 0& 0 & 0\\
0 & 0 & V & 0 & 0 & 0& 0 & 0 & 0 & 0 & 0 & 0& 0 & 0\\
0 & 0 & 0 & V & 0 & 0 & 0 & 0 & 0 & 0& 0 & 0& 0 & 0\\
0 & 0 & 0 & 0 & V & 0& 0 & 0 & 0 & 0 & 0 & 0& 0 & 0\\
0 & 0 & 0 & 0 & 0 & V& 0 & 0 & 0 & 0 & 0 & 0& 0 & 0\\
0 & 0 & 0 & 0 & 0 & 0& V & 0 & 0 & 0 & 0 & 0& 0 & 0\\
0 & 0 & 0 & 0 & 0 & 0& 0 & V^{3} & 0 & 0& 0 & 0 & 0 & 0\\
0 & 0 & 0 & 0 & 0 & 0& 0 & 0 & V^{-1} & 0 & 0 & 0& 0 & 0\\
0 & 0 & 0 & 0 & 0 & 0& 0 & 0 & 0 & V^{-1} & 0 & 0& 0 & 0\\
0 & 0 & 0 & 0 & 0 & 0& 0 & 0 & 0 & 0 & V^{-1} & 0& 0 & 0\\
0 & 0 & 0 & 0 & 0 & 0& 0 & 0 & 0 & 0 & 0 & V^{-1}& 0 & 0 \\
0 & 0 & 0 & 0 & 0 & 0& 0 & 0 & 0 & 0 & 0 & 0& V^{-1} & 0\\
0 & 0 & 0 & 0 & 0 & 0& 0 & 0 & 0 & 0 & 0 & 0& 0 & V^{-1} 
\end{array}\right).
\end{equation}


\section*{Acknowledgments}

 The  work of CH is supported by the EPSRC Programme
Grant EP/K034456/1,
and by the STFC
Consolidated Grant ST/L00044X/1.

\appendix
\section{Appendix: T-fold solutions}
The metric and $B$-field of the T-fold which is T-dual to the $T^2$ bundle over $T^4$ fibred over a line
is given by T-dualising the metric in the $z^3$ direction. This results in:
\begin{equation}
g =     \left(\begin{array}{cccccc}A_5  & -C& 0 &  m  z^6 C& m  z^6 A_5& 0
\\
 -C& A_4& 0
  & 
  -m  z^6 A_4
  & -m  z^6 C
& 0
\\ 0 & 0 & \frac{1}{f}& 0 & 0 & 0
\\ m  z^6 C&
-m  z^6 A_4
& 0 & m^2\,{(z^6)}^2
-D_5
+1 & m^2 ( z^6)^2 C
& 0
\\
m  z^6 A_5
 &
  -m  z^6 C
   & 0 
   & m^2 ( z^6)^2 C
   &
    m^2\,{(z^6)}^2-
   D_4
    +1 & 0
    \\ 0 & 0 & 0 & 0 & 0 & 1 \end{array}\right),\label{T-fold1metric}
\end{equation}
\begin{equation}
B =     \left(\begin{array}{cccccc} 0 & 0 & \frac{m\,z^4}{f}& 0 & 0 & 0\\ 0 & 0 & \frac{m\,z^5}{f}& 0 & 0 & 0\\ -\frac{m\,z^4}{f}& -\frac{m\,z^5}{f}& 0 & \frac{m^2\,z^5\,z^6}{f}& -\frac{m^2\,z^4\,z^6}{f}& 0\\ 0 & 0 & -\frac{m^2\,z^5\,z^6}{f}& 0 & 0 & 0\\ 0 & 0 & \frac{m^2\,z^4\,z^6}{f}& 0 & 0 & 0\\ 0 & 0 & 0 & 0 & 0 & 0 \end{array}\right),\label{T-fold1Bfield}
\end{equation}
where 
$$f = {1+m^2\,{(z^4)}^2+m^2\,{(z^5)}^2} $$
and
\begin{equation}
A_5=\frac{m^2\,{(z^5)}^2+1}{f}, \qquad A_4=\frac{m^2\,{(z^4)}^2+1}{f}, \qquad C=\frac{m^2\,z^4\,z^5}{f}
\end{equation}
\begin{equation}
D_5=\frac{m^4\,{(z^5)}^2\,{(z^6)}^2}{f}
,\qquad
D_4= \frac{m^4\,{(z^4)}^2\,{(z^6)}^2}{f}
\end{equation}

The metric and $B$-field of the T-fold which is T-dual to the $T^3$ bundle over $T^4$ fibred over a line
is given by T-dualising the metric in the $z^4$ direction. This results in:
\begin{equation}
g =     \left(\begin{array}{ccccccc} A_{5} & -C_{56} & -C_{57} & 0 & 0 & m\,z^7-D_{557} & 0\\ -C_{56} & A_{6} & -C_{67} & 0 & -m\,z^7 & -D_{567} & 0\\ 
-C_{57}& -C_{67} & A_{7} & 0 & m\,z^6 & -D_{577}& 0\\ 
0 & 0 & 0 & \frac{1}{f} & 0 & 0 & 0\\ 
0 & -m\,z^7 & m\,z^6 & 0 & m^2\,{(z^6)}^2+m^2\,{(z^7)}^2+1 & 0 & 0\\ 
m\,z^7-D_{557} & -D_{567} & -D_{577} & 0 & 0 & m^2\,{(z^7)}^2-\frac{m^4\,{(z^5)}^2\,{(z^7)}^2}{f}+1 & 0\\ 
0 & 0 & 0 & 0 & 0 & 0 & 1 \end{array}\right),\label{T-fold2metric}
\end{equation}
\begin{equation}
B =     \left(\begin{array}{ccccccc} 0 & 0 & 0 & \frac{m\,z^5}{f} & 0 & 0 & 0\\ 0 & 0 & 0 & \frac{m\,z^6}{f} & 0 & 0 & 0\\ 0 & 0 & 0 & \frac{m\,z^7}{f} & 0 & 0 & 0\\ -\frac{m\,z^5}{f} & -\frac{m\,z^6}{f} & -\frac{m\,z^7}{f} & 0 & 0 & -\frac{m^2\,z^5\,z^7}{f} & 0\\ 0 & 0 & 0 & 0 & 0 & 0 & 0\\ 0 & 0 & 0 & \frac{m^2\,z^5\,z^7}{f} & 0 & 0 & 0\\ 0 & 0 & 0 & 0 & 0 & 0 & 0 \end{array}\right),\label{T-fold2Bfield}
\end{equation}
where $$ f ={1 + m^2\,{(z^5)}^2+m^2\,{(z^6)}^2+m^2\,{(z^7)}^2},$$
$$A_5 = 1-\frac{m^2\,{(z^5)}^2}{f} , A_6 = 1-\frac{m^2\,{(z^6)}^2}{f} , A_7 = 1-\frac{m^2\,{(z^7)}^2}{f},$$
$$C_{56} = \frac{m^2\,z^5\,z^6}{f} , C_{57} = \frac{m^2\,z^5\,z^7}{f}, C_{67} = \frac{m^2\,z^6\,z^7}{f}, $$

$$D_{557} = \frac{m^3\,{(z^5)}^2\,z^7}{f}, D_{567} = \frac{m^3\,z^5\,z^6\,z^7}{f}, D_{577} = \frac{m^3\,(z^5)\,{(z^7)}^2}{f}  $$




\end{document}